\documentclass[twocolumn,traditabstract]{aa} 
\usepackage{graphicx}
\usepackage{psfrag}
\usepackage{amsmath}
\usepackage{amssymb}
\usepackage{color,xcolor}
\usepackage{hyperref}
\usepackage{caption,natbib}
\hypersetup{
    colorlinks=true,
    citecolor=blue,
    filecolor=black,
    linkcolor=blue,
    urlcolor=magenta,
    linktocpage=true,
    breaklinks = true
}
\usepackage{hhline}
\usepackage{enumitem}
\usepackage{subfigure,float}
\usepackage{placeins}
\usepackage{multirow}
\usepackage{makecell}
\setcellgapes{6pt}
\makegapedcells
\usepackage[normalem]{ulem}
\usepackage{dashrule}
\usepackage{array,booktabs}
\usepackage{siunitx}
\DeclareUnicodeCharacter{2212}{\textendash}

\newcommand{\beq}{\begin{equation}}
\newcommand{\eeq}{\end{equation}}

\newcommand{\ber}{\begin{eqnarray}}
\newcommand{\eer}{\end{eqnarray}}

\newcommand{\ba}{\begin{align}}
\newcommand{\ea}{\end{align}}

\def \Ne {N_\text{obs,m-b}}

\def \vecx {\mathbf{x}}
\def \vecu {\mathbf{u}}
\def \vecv {\mathbf{v}}
\def \var {\mathrm{Var}}
\def \Ix {I(\vecx)}
\def \varp {\mathrm{Var}_{\mathrm{P}}}
\def \varo {\mathrm{Var}_{\mathrm{bg}}}

\newcommand{\ord}[1]{\!\times\!10^{#1}}

\begin{document}

   \title{HOLISMOKES} \subtitle{XXI. Detecting strongly lensed type Ia supernovae from time series of multi-band LSST-like imaging data -- Part II}

\author{Satadru Bag\inst{1,2}\thanks{satadru.bag@tum.de}, Raoul Ca{\~n}ameras\inst{3}, Sherry H.~Suyu\inst{1,2}, Stefan Schuldt\inst{4,5,6}, Stefan Taubenberger \inst{1,2}, Irham Taufik Andika\inst{7,1,2}, Alejandra Melo\inst{8,1,2}, 
Ming Kei Chan\inst{9,1}
          }

   \institute{Technical University of Munich, TUM School of Natural Sciences, Physics Department,  James-Franck-Stra{\ss}e 1, 85748 Garching, Germany
         \and 
           Max-Planck-Institut f{\"u}r Astrophysik, Karl-Schwarzschild Stra{\ss}e 1, 85748 Garching, Germany
           \and
          Aix Marseille Univ, CNRS, CNES, LAM, Marseille, France
           \and
          Finnish Centre for Astronomy with ESO (FINCA), University of Turku, FI-20014 Turku, Finland
          \and
          Department of Physics, P.O. Box 64, University of Helsinki, FI-00014
Helsinki, Finland
          \and
          INAF - IASF Milano, via A. Corti 12, I-20133 Milano, Italy
          \and
          University Observatory, LMU Faculty of Physics, Scheinerstrasse 1, 81679 Munich, Germany
          \and
          European Southern Observatory, Karl-Schwarzschild Stra{\ss}e 2, 85748 Garching, Germany
          \and
          Hong Kong University of Science and Technology, Clear Water Bay, Kowloon, Hong Kong SAR
}



  \abstract
{
Strong gravitationally lensed supernovae (LSNe) are rare but extremely valuable probes of cosmology and astrophysics. Identifying them promptly within the massive alert streams of time-domain surveys such as the Rubin Legacy Survey of Space and Time (LSST) is essential for enabling timely follow-up observations. In our previous study, Bag et al. (2026), we introduced a deep-learning framework for detecting LSNe Ia directly from multi-band, multi-epoch image cutouts. The model employs a convolutional long short-term memory (\texttt{ConvLSTM}) architecture to capture spatial and temporal correlations in time-series imaging data, enabling classification to be updated as new observations arrive. In this work, we extend that framework by incorporating greater astrophysical and observational realism into the simulations. In particular, we present a method to construct realistic image time series from single-epoch observations by introducing epoch-to-epoch point spread function variations while consistently correcting the corresponding variance maps. The dataset is constructed using Hyper Suprime-Cam (HSC) PDR3 observations and includes simulated lensed host-galaxy arcs, realistic SN light-curve variations, and Poisson noise. We also introduce an additional negative class consisting of SN Ia occurring in the foreground lens galaxy, representing an astrophysically motivated and challenging source of false positives. Despite these additional complexities, the model retains strong performance. The receiver operating characteristic improves rapidly during the first few observations, reaching a true-positive rate of $\sim60\%$ at a false-positive rate of $\mathcal{O}(10^{-4})$ by the seventh observation and $\sim80\%$ by the tenth. Detectable host-galaxy arcs provide additional spatial information that improves early performance, although the detection efficiency remains broadly consistent across different SN-host offsets. We also investigate potential confusion with sibling SNe occurring in luminous red galaxies and identify the configurations that most closely mimic lensed systems. These results demonstrate that the image-time-series approach remains robust under more realistic observing conditions, and is well suited for real-time LSN searches in LSST and other time-domain surveys.
}
   {}
   {}
   {}
   {}


   \titlerunning{Detecting lensed Type Ia supernovae in time series of multi-band LSST-like imaging data -- Part II}
   \authorrunning{Bag et al}

   \maketitle

\section{Introduction}

Strong gravitational lensing occurs when the gravitational potential of a massive galaxy or galaxy cluster bends light from a background source, producing multiple magnified images and, for extended sources, characteristic arcs and distortions (see, e.g., \cite{1996astro.ph..6001N} for a pedagogical introduction). Acting as a cosmic telescope, strong lensing enables a wide range of astrophysical and cosmological studies, including galaxy evolution, the stellar initial mass function, and the nature of dark matter \citep[e.g.,][]{Mao:1997ek, Metcalf:2001ap, Dalal2002, Pooley_2009, Oguri2014, Treu2010, Jim_nez_Vicente_2019,shajib2024,vegetti2023}. One of its most powerful applications is time-delay cosmography (TDC), which uses time delays between multiple images of variable background sources to measure cosmological distance ratios and constrain the Hubble constant independently of the local distance ladder \citep[e.g.,][]{Refsdal1964_1, Refsdal1964_2, Saha, Oguri_2007, 2016A&ARv..24...11T, Bonvin2017, Wong:2019kwg, 2020A&A...643A.165B, Birrer:2020jyr, Treu:2022aqp, Suyu_2024, TDCOSMO2025}. These measurements provide an important avenue for addressing the persistent $H_0$ tension \citep{Planck:2018vyg, Riess2022,valentino2021,verde2024} and, when multiple sources at different redshifts are available, can also probe dark energy and spatial curvature \citep{Grillo2024, Linder2011, Jee2016, Bolamperti24}.

TDC requires time-variable background sources, most notably quasars (QSOs) and supernovae (SNe) \citep{Treu2010}. While lensed quasars have historically formed the backbone of such studies, lensed supernovae (LSNe) offer several advantages \citep{Suyu_2024, LSNe_rev2}. Their well-defined light curves, particularly for Type Ia SNe, help reduce modeling degeneracies and enable more precise time-delay measurements. LSNe also provide a unique opportunity to study SN progenitors, as detection of the first image allows lens modeling to predict subsequent images and enable very early follow-up observations \citep{Suyu20}.

LSNe offer substantial scientific value, but remain extremely rare, with only a handful discovered to date \citep{Kelly:2014mwa,Goobar:2016uuf,2021NatAs.tmp..164R,chen2022,Goobar2023,Frye2023,sn_encore,SNwny1,SNwny2, Coulter2026,lemon26}. Upcoming wide-field time-domain surveys such as the Legacy Survey of Space and Time (LSST) at the Vera C. Rubin Observatory \citep{lsst1,ivezic19} and the Nancy Grace Roman Space Telescope \citep{roman2015, Pierel2021} are expected to dramatically increase their discovery rate. LSST alone is predicted to detect hundreds of LSNe over its ten-year survey, along with thousands of lensed quasars \citep{om10,Ana_2023,Nikki_2023,Bag:2024kbk}.

Strongly lensed systems have already demonstrated their power for cosmography. The H0LiCOW collaboration measured $H_0$ with $\sim2.4\%$ precision using six lensed quasars, although individual systems typically constrain $H_0$ at the $5$--$10\%$ level \citep{Wong:2019kwg}. Relaxing assumptions on the lens mass density profile increases the uncertainty, with recent TDCOSMO analyses yielding $\sim5\%$ precision from eight systems \citep{TDCOSMO2025}. Achieving percent-level precision with LSNe therefore requires discovering tens to hundreds of new systems \citep{Suyu20,Nikki_2023}. Realizing this potential, however, requires rapid and reliable identification of LSNe, ideally before peak brightness to enable timely follow-up. This poses a major challenge for LSST, which is expected to generate $\sim10^7$ alerts per night, rendering manual inspection impractical.

Several strategies have been developed to identify LSNe, including direct detection of multiple images \citep{om10}, magnification-based selection using standardized Type Ia luminosities \citep[e.g.,][]{goobar17,goldstein19,sagues24}, color--magnitude cuts \citep{Quimby2014}, and light-curve-based methods for unresolved systems \citep{Bag:2020pbg,misha2021,misha2022,Bag:2024kbk,Elahe2025, Park:2026hmc}. Another fundamentally different approach relies on cross-matching transient alerts with catalogs of known strong lenses \citep[e.g.,][]{shu18,craig24}, enabled by recent deep-learning searches that have produced large samples of galaxy- and cluster-scale strong lenses in wide-field surveys \citep[e.g.,][]{petrillo17,jacobs19,metcalf19,canameras20,canameras21,schuldt25, Euclid_Q1}. While effective in certain regimes, these methods often depend on resolving multiple images, detecting host galaxy arcs, or extracting reliable light curves, which becomes increasingly difficult for small-separation systems and during the early phases of the explosion. Rapid identification is nonetheless essential to enable early mass modeling and coordinated follow-up, particularly for programs such as HOLISMOKES \citep{Suyu20} that aims to constrain $H_0$ and SN progenitors. Given the $\mathcal{O}(10^7)$ nightly alerts expected from LSST, automated identification through machine-learning approaches becomes unavoidable.

In HOLISMOKES Paper XVII \citep{Bag2025}, we presented a deep-learning approach that operates directly on multi-band image sequences to identify LSNe. Previous studies have largely relied on pipelines combining convolutional neural networks (CNNs) applied to reference or difference images with recurrent architectures such as Long Short-Term Memory (\texttt{LSTM}) networks operating on extracted light curves \citep{Morgan2022, Morgan2023}. Instead, we developed a framework based on a Convolutional Long Short-Term Memory (\texttt{ConvLSTM}) architecture \citep{convlstm}, which processes multi-band time series of image cutouts and captures spatial and temporal correlations simultaneously. Similar \texttt{ConvLSTM}-based approaches have been explored using fully simulated single-band datasets \citep{Ramanah2022}. Our implementation differs in that it operates on multi-band sequences and is trained on realistic simulations constructed from real survey data.

The model in HOLISMOKES XVII was trained and validated on simulations generated by ingesting real Hyper Suprime-Cam (HSC) observations \citep{aihara18, yasuda19}, closely matching LSST depth and filter characteristics. Lensed and unlensed SNe were injected into observed HSC images of luminous red galaxies (LRGs), preserving realistic backgrounds, noise properties, point spread functions (PSFs), and image artifacts. In that work we focused on Type Ia LSNe without detectable host-galaxy light, providing a proof of concept that early identification of LSNe is feasible using only a small number of observations distributed across different bands.

In this paper, we build upon HOLISMOKES XVII and extend the methodology to address several of its simplifying assumptions. First, we expand the training and validation datasets using the HSC Public Data Release~3 (PDR3), increasing the sample size by approximately $50\%$ relative to the previous PDR2-based dataset. We also incorporate lensed host-galaxy arcs into the simulated LSNe. While hostless systems remain an important discovery channel, a substantial fraction of LSNe are expected to exhibit detectable host arcs in LSST \citep{Ana_2024}; including these features therefore broadens the applicability of the model.

Second, we introduce additional realism in both the transient and observational modeling. The PSF and the corresponding Poisson noise realization are allowed to vary across epochs in all simulated time series, reflecting changing observing conditions in time-domain surveys. We also adopt SALT2-based Type Ia SN light curves that incorporate intrinsic variations in stretch and color. Finally, we introduce a new astrophysically motivated negative class consisting of SNe occurring in the foreground lens galaxy itself. Although these events are not lensed, they are expected to represent a dominant source of false positives in LSNe searches, providing a more stringent and realistic test of the model's discriminative power.

The remainder of this paper is organized as follows. Section~\ref{sec:sim_data_method} summarizes the data acquisition and simulation methodology, highlighting the main improvements relative to HOLISMOKES XVII and referring there when procedures remain unchanged. Section~\ref{sec:model} describes the \texttt{ConvLSTM}-based model architecture and data normalization scheme, which largely follow HOLISMOKES XVII. The classification performance and robustness tests are presented in Section~\ref{sec:results}. Finally, Section~\ref{sec:conclusions} summarizes our conclusions and discusses the implications for LSNe searches in the LSST-era.

Throughout this study we use image cutouts of $59\times59$ pixels, corresponding to a field of view of approximately $10\arcsec \times 10\arcsec$. We assume a flat $\Lambda$CDM cosmology with $\Omega_{\rm M}=0.308$ \citep{planck16}, and adopt $H_0=72\,{\rm km\,s^{-1}\,Mpc^{-1}}$ \citep{Bonvin2017}.

\section{Simulation methodology}
\label{sec:sim_data_method}
\begin{table*}[h]
\centering
\renewcommand{\arraystretch}{1.3}
\begin{tabular}{|c|c|c||c|c|c|c|}
\cline{3-7}
\multicolumn{2}{c|}{} & Positive & \multicolumn{4}{c|}{Negative} \\
\cline{3-7}
 \multicolumn{2}{c|}{} &  \makecell{LSNe Ia \\ (Set $P$)} & \makecell{HSC variables \\ (Set $N_1$)} & \makecell{SN Ia in lens \\ (Set $N_2$)} & \makecell{SN Ia in LRGs\\ (Set $N_3$)} & \makecell{SN Ia in spirals\\ (Set $N_4$)} \\
\hline
\multicolumn{2}{|c|}{\makecell{Ingested \\ observed \\ multi-band \\ HSC data}} & \makecell{Co-added \\ LRGs from \\ the wide-layer \\ $\Rightarrow$ lens; \\ HUDF galaxies \\ $\Rightarrow$ SN hosts}  & 
\makecell{Variables in the \\ COSMOS field \\ from the HTS at \\ different epochs \\ with `suitable' \\cadence and depths}& 
\makecell{Co-added \\ LRGs from \\ the wide-layer \\ $\Rightarrow$ lens; \\ HUDF galaxies \\ $\Rightarrow$ source galaxies} & \makecell{Co-added \\ LRGs from \\ the wide-layer\\ $\Rightarrow$ SN Ia hosts}  & \makecell{Co-added spiral \\ galaxies from \\ the wide-layer\\ $\Rightarrow$ SN Ia hosts} \\
\hline
\multicolumn{2}{|c|}{\makecell{Injected at \\ different epochs}} & \makecell{LSNe Ia images \\ \& host arcs } & \makecell{Fully observed, \\ no simulation} & \makecell{SNe Ia in LRGs \\ \& lensed host arcs}  & \makecell{SNe Ia \\ in LRGs} & \makecell{SNe Ia in \\ spiral galaxies} \\
\hline

\multicolumn{2}{|c|}{\makecell{Relevant sec. of \\ \citet{Bag2025}}} &  \makecell{2.2, \\ Appendix A} & 2.4 & Newly added  &  2.5.1& 2.5.2 \\
\hline
\noalign{\vskip 0.6pt}
\hline

\multirow{2}{*}{\rotatebox{90}{Number~~}} 
 & \makecell{Training \\ ($128\ord{3}$ total)} & $64\ord{3} $ & $12\ord{3} ~  (18.75\%)$ & $20\ord{3} ~  (31.25\%)$ &  $20\ord{3} ~  (31.25\%)$ & $12\ord{3} ~  (18.75\%)$\\
\cline{2-7}
 & \makecell{Validation + Test\\($104.6\ord{3}$ total) } & $8.6\ord{3}$ & $32\ord{3} ~  (33.33\%)$ & $16\ord{3} ~  (16.67\%)$ &  $16\ord{3} ~  (16.67\%)$ & $32\ord{3} ~  (33.33\%)$\\
\hline
\end{tabular}
\caption{
Summary of the positive and negative classes used to construct the training, validation, and test datasets. The top section describes the observational inputs, their role in the simulations, and references to the corresponding sections of HOLISMOKES XVII \citep{Bag2025}. The positive class consists of simulated LSNe~Ia systems, while the negative classes include HSC variables and three subclasses of normal SNe~Ia occurring in different environments: foreground lens galaxies in the lensing systems, LRGs, and spiral galaxies. 
The bottom section lists the number of samples used in the training and validation+test sets. The percentages in parentheses for the negative subclasses indicate their fractional contribution to the total negative sample. The subclass composition of the validation and test sets differs from that of the training set, where $N_2$ and $N_3$ are oversampled in anticipation that they will be the most challenging to distinguish from the positive LSNe~Ia cases.}
\label{tab:details}
\end{table*}

\begin{figure*}
    \centering
    \includegraphics[width=\textwidth]{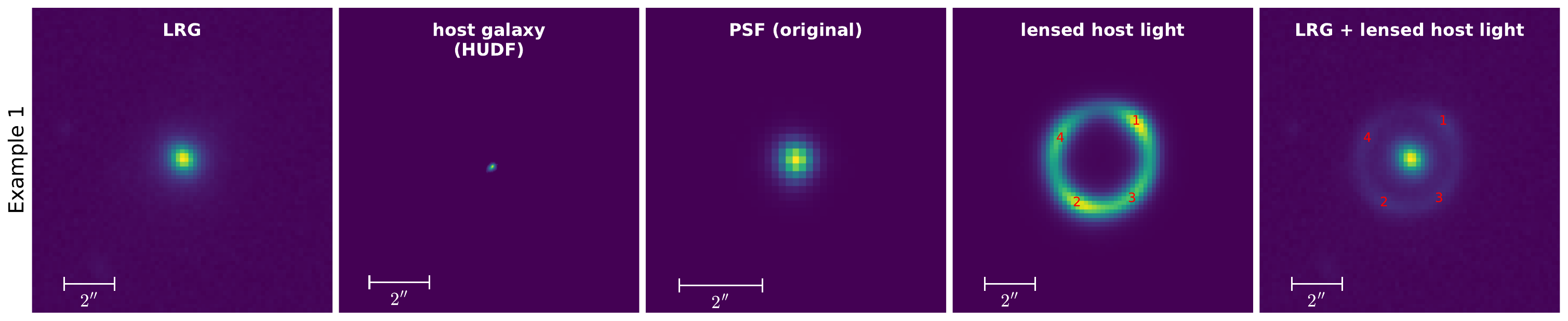}\\
    \vspace*{-2mm}
    \includegraphics[width=\textwidth]{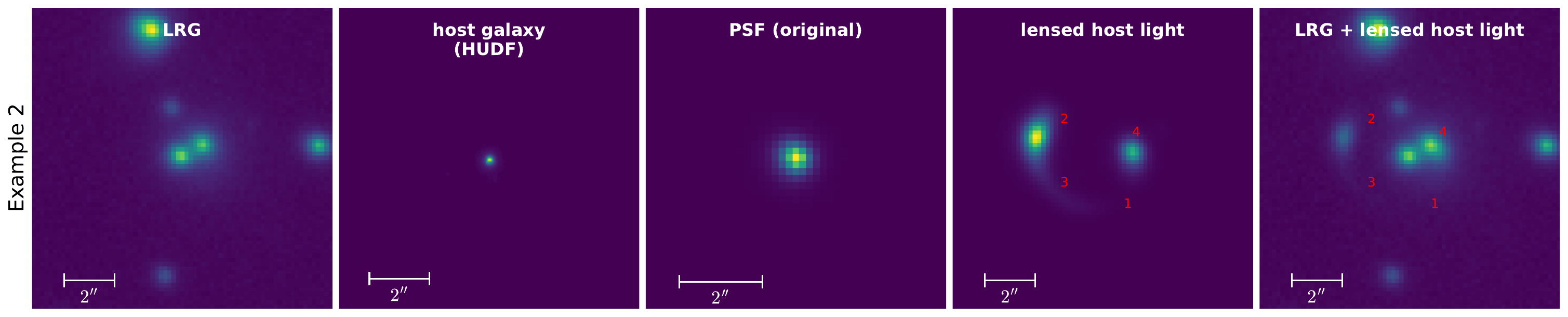}
\caption{
Two representative examples of LSNe Ia are shown in the two rows using $i$-band images. For each example, the panels (from left to right) display the cutouts of the LRG lens, the HUDF galaxy used as the host, the PSF, the PSF-convolved lensed host light at HSC resolution, and the combined LRG + lensed host light. The host galaxy image is shown at the native HUDF resolution ($0.03\arcsec$), whereas the other panels are displayed at the reference HSC resolution ($0.168\arcsec$). In the last two panels we mark the positions of the LSNe images, which form quads in both examples (labelled by 1, 2, 3, 4 in red). In the bottom example, the SN lies within the diamond caustic and is therefore quadruply imaged, while the host galaxy lies outside the caustic and is only doubly imaged. This example also contains several foreground and background objects, making the configuration more challenging for visual identification.
}
    \label{fig:host_light_demo}
\end{figure*}

\begin{figure*}[h!]
\centering
\includegraphics[width=\textwidth]{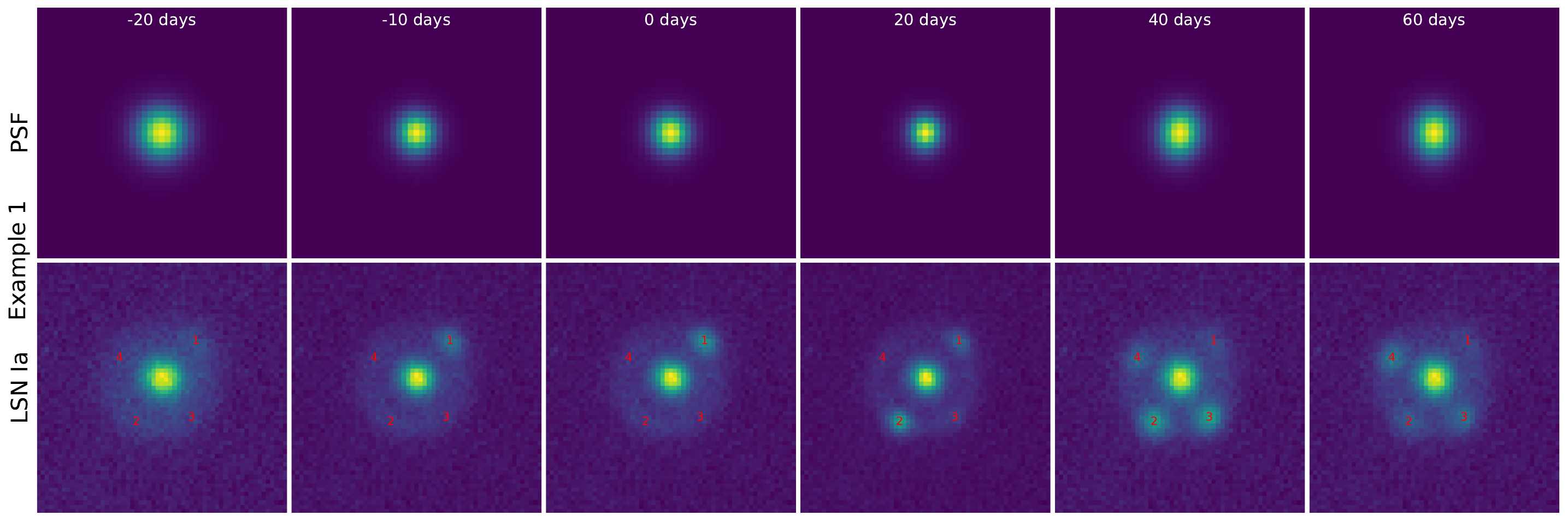} \\
\vspace*{-2 mm}
\includegraphics[width=\textwidth]{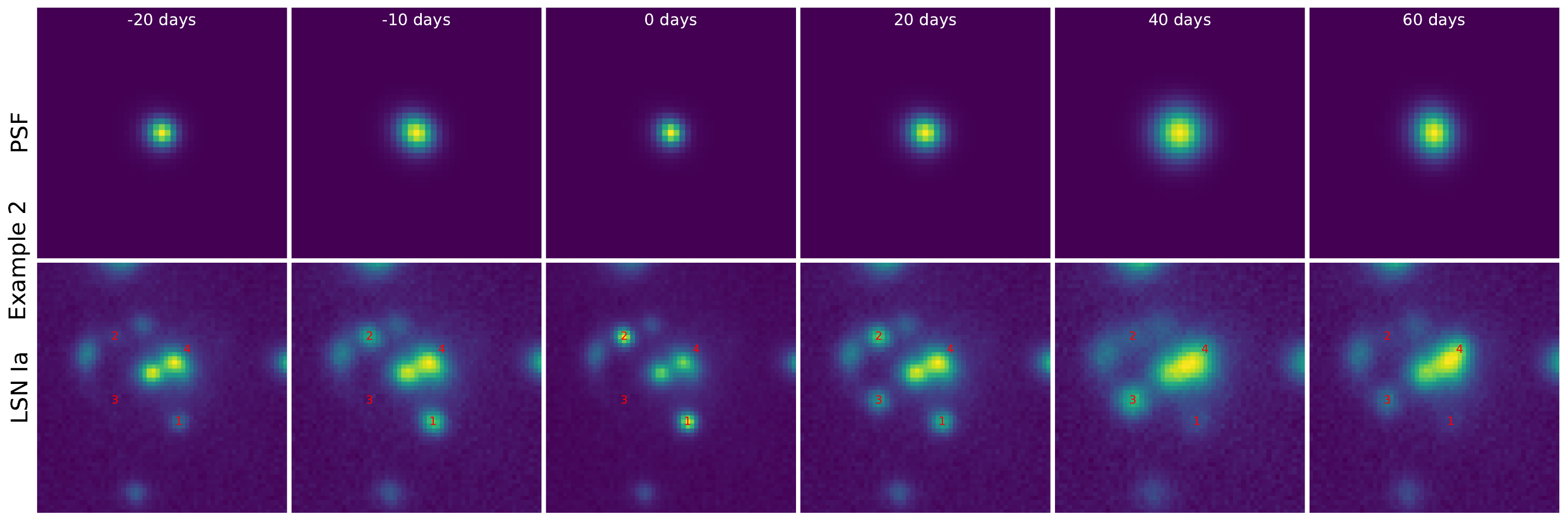}\\
\vspace*{-2 mm}
\includegraphics[width=\textwidth]{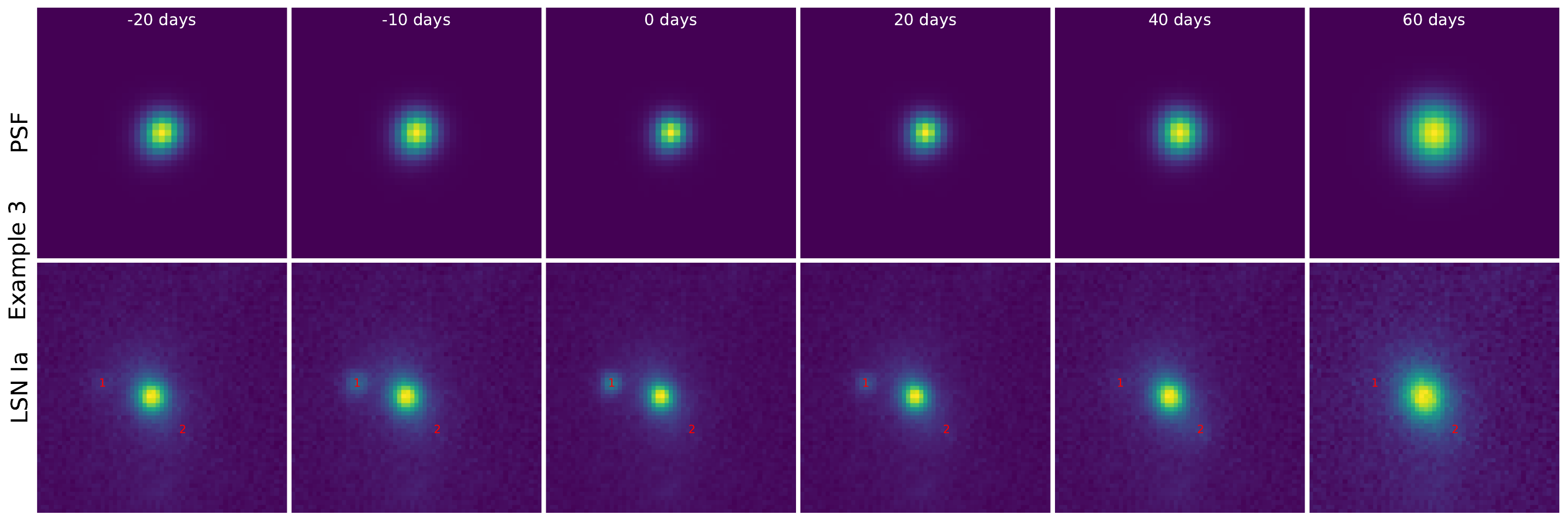}\\
\vspace*{-2 mm}
\caption{
Time series of LSNe Ia illustrating the effect of varying PSF conditions across epochs. Three representative examples are shown, each represented by two consecutive rows. The upper row shows the PSF at different observing epochs, with the corresponding observation time indicated, while the lower row shows the corresponding images of the LSNe Ia system. The first two examples (top four rows) correspond to the systems shown in Fig.~\ref{fig:host_light_demo}. The third example (bottom two rows) illustrates a configuration in which the lensed host galaxy is not detectable at the survey depth, leaving only the multiple SN images visible in the time series. The temporal evolution of the SN brightness follows the light curves presented in Fig.~\ref{fig:lc_demo}, including the effects of lensing magnification and time delays between the multiple SN images. All panels are shown in the $i$-band for demonstration. The time axis is defined such that 0 days corresponds to the peak phase of the first-arriving SN image for visual clarity. In the actual training data, however, the temporal reference is defined by setting the first detection to 0 days for all samples.
}
\label{fig:ts_demo}
\end{figure*}

\begin{figure*}
\centering
\includegraphics[width=\textwidth]{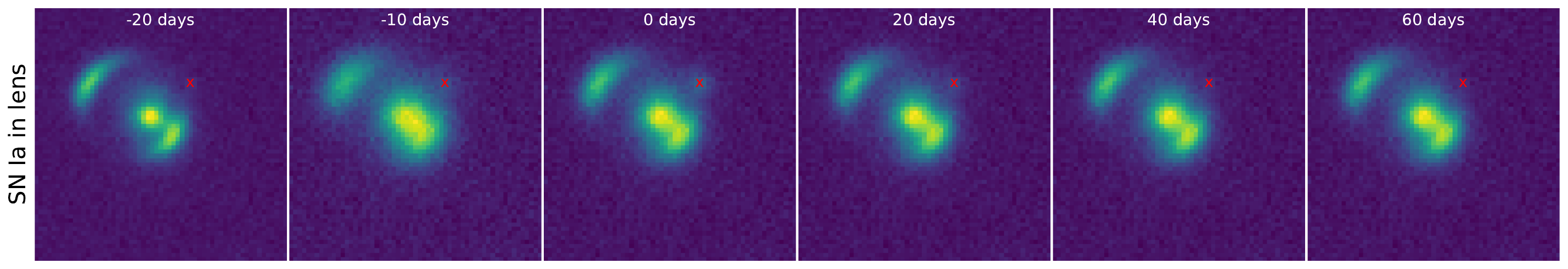}\\
\vspace*{-2mm}
\includegraphics[width=\textwidth]{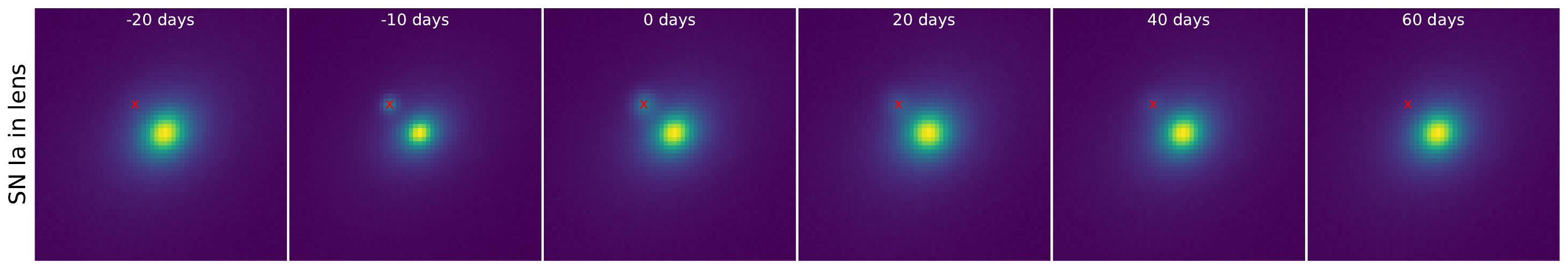}
\caption{
Two examples of SN Ia occurring in the foreground lens galaxy of a 
lensing system, representing the ``SN in lenses'' subclass (Set $N_2$) 
considered in this work as one of the most challenging sources of false 
positives. In the top row, the lensed background host galaxy is clearly 
identifiable. In contrast, the bottom row shows a system where the 
background galaxy remains undetectable at the survey 
depth. The PSF and Poisson noise vary across the time series, similar to the 
examples shown in Fig.~\ref{fig:ts_demo}. Time is measured relative to the $i$-band peak of the SN Ia. For clarity of illustration, we show examples where the SN, marked by red crosses, 
is located relatively far from the LRG center.}
\label{fig:sn_in_lenses_demo}
\end{figure*}

In the absence of sufficient real lensed transient data, we rely on simulations constructed to be as realistic as possible by ingesting real HSC observations, thereby preserving observational complexities such as sky background, unrelated foreground and background sources, saturation effects, and detector artifacts.

\subsection{Summary of the datasets}
 The overall procedure for selecting these observations (see Appendix \ref{app:ingredient_details} for a brief note on this) and incorporating them into the training dataset follows the framework introduced in HOLISMOKES XVII \citep{Bag2025}. One notable change relative to that work is the transition from the HSC public data release 2 (PDR2) to PDR3, which significantly increases the size of the parent sample. The main methodological improvements to the framework are described in subsections \ref{sec:host_add} - \ref{sec:satl2_imp} below. We focus on Type Ia SNe (hereafter SNe Ia) in this work also; however, the analysis framework is sufficiently general to be extended to other types of transients.

We formulate the identification of LSNe Ia as a binary classification problem. The positive class (Set $P$) consists of simulated galaxy-scale LSNe Ia systems constructed by injecting point-like SN images into HSC Wide-layer observations of luminous red galaxies (LRGs). The lensing configurations are generated using our \texttt{GLEE}-based \citep{glee1, glee2} simulation pipeline \citep{schuldt21, schuldt23}. LRGs are adopted as foreground lenses because of their high lensing efficiency, and details of the LRG sample selection are provided in Appendix \ref{app:LRG_selection}. From a parent sample of $119,000$ LRGs we construct $113,000$ lens--source pairs following the OM10 \citep{om10, Oguri2018} redshift distributions and assuming a flat Einstein radius distribution between $0.1$ and $2.0\arcsec$. In this work, however, we restrict the sample to systems with $\theta_{\rm E} > 0.5\arcsec$, since our adopted seeing distribution (explained in Sec. \ref{sec:seeing_sampling_psf}) already yields a sufficient number of unresolved LSNe toward the lower end of this range. SN Ia images are injected only for configurations that produce detectable lensed images ($\geq 5\sigma$) in our simulations, with the SN placed at the source redshift.

The negative class represents realistic contaminants in LSNe searches and includes:
\begin{itemize}
    \item Set $N_1$: variable sources selected from the HSC transient survey (HTS), see Appendix~\ref{app:HTS} for brief details.
    \item Set $N_2$: a new subclass of SNe~Ia occurring in the foreground lens galaxy within a lensing system and therefore not lensed (described in Sec.~\ref{sec:sn_in_lens}). We anticipate these to be among the most difficult cases to distinguish from the positive LSNe systems.
    \item Sets $N_3$ and $N_4$: normal SNe~Ia in non-lens galaxies, hosted by LRGs and spiral galaxies, respectively.
\end{itemize}
While placing normal SNe Ia in Sets $N_2$, $N_3$, and $N_4$, we draw the projected offset between the SN and the galaxy center from a uniform distribution $\mathcal{U}(0,2R_{50})$ with a random orientation angle, where $R_{50}$ denotes the half-light radius of the host galaxy. The offset is further clipped at $2\arcsec$. Astrophysically motivated priors for the SN-host offset, based on stellar or light-density
profiles, have been investigated \citep{Lokken_2023}, but their detailed dependence on galaxy
type, redshift, and other properties remains uncertain. We therefore adopt a uniform
distribution for the SN-host center offset in the training set to ensure equal
representation across configurations, the same choice adopted in \citet{Bag2025}. 

The simulation setup and the composition of the datasets used for training, validation, and testing are summarized in Table~\ref{tab:details}. The top portion outlines the observational inputs used in each subclass simulation, their role in the pipeline, and references to the corresponding descriptions in HOLISMOKES XVII. The bottom portion lists the number of samples used in the training and validation+test datasets. The validation and test sets follow a different subclass composition than the training set and together contain $104,600$ samples, of which $16,000$ are used for validation and the remainder for testing. Sets $N_2$ and $N_3$, expected to be the most difficult to distinguish from the LSNe~Ia cases, are oversampled in the training set.

In the following subsections \ref{sec:host_add}--\ref{sec:satl2_imp} we describe the improvements introduced in this work relative to HOLISMOKES XVII, which increase the astrophysical and observational realism of the training set.

\subsection{Incorporation of lensed host galaxies}
\label{sec:host_add}

We incorporate real galaxies observed in the Hubble Ultra Deep Field \citep[HUDF;][]{beckwith06} as host galaxies for LSNe Ia. Modeling realistic hosts up to high-redshifts requires high-resolution and high signal-to-noise (S/N) multi-band imaging motivating the use of HST/ACS cutouts from the HUDF with a pixel scale of approximately $0.03\arcsec$.

Our sample consists of about 1400 HUDF galaxies with spectroscopic redshift measurements from MUSE \citep{inami17}, covering up to z$\simeq$3, and with neighboring galaxies masked out from the ACS cutouts following \citet{schuldt21, schuldt23}. For each of the 113,000 lens-source pairs described above, we randomly select a host galaxy from the HUDF sample within the corresponding redshift bin, allowing individual galaxies to be reused multiple times. For each lens-source pair, once a viable lensing configuration is identified, a HUDF galaxy is assigned as the host of the LSNe Ia. Finally, color corrections are applied to transform the HST/ACS filter passbands (F435W, F606W, F775W, and F850LP) to match the HSC zeropoints in $griz$ bands.

Several studies \citep[e.g.,][]{Ana_2024} indicate that roughly half of the LSNe~Ia detectable by LSST are expected to have host galaxies that are themselves lensed and detectable. In our mock LSNe~Ia systems, the SN is always required to be lensed and sufficiently magnified to be detected, whereas the detectability of the host galaxy depends on both its brightness and the offset between the SN and the host-galaxy center.

For each system, the center of the assigned HUDF host galaxy is placed on the source plane at a distance drawn from $\mathcal{U}(0,3R_{50})$ from the SN position, with a random orientation. The offset is clipped at $2\arcsec$, as for the offsets used for normal SNe~Ia in the negative sets. Compared to the negative sets, we adopt a slightly wider offset range for the lensed systems to better sample configurations that produce diverse lensing morphologies, including cases where the SN and the host galaxy cross caustics differently and form different numbers or arrangements of images.

The lensed host-galaxy arcs are computed on the image plane using our lens simulation pipeline \citep{schuldt21, schuldt23}. The procedure is illustrated in Fig.~\ref{fig:host_light_demo}, which shows two example systems in two rows. For each system, the assigned HUDF galaxy is placed at the source redshift and ray-traced through the lens model to generate the lensed host-light distribution. The resulting image, initially at the native HUDF resolution ($0.03\arcsec$), is resampled to the HSC resolution ($0.168\arcsec$), convolved with the appropriate PSF, and combined with the foreground LRG cutout. The lensed host-galaxy light remains fixed across epochs, apart from variations introduced by the epoch-dependent PSF and Poisson noise, whereas the LSNe~Ia brightness evolves according to its intrinsic light curve, lensing magnification, and time delays. The positions of the multiple SN images are determined from the same lens model, accounting for the offset between the SN and the host-galaxy center, and are marked by the red labels ($1$--$4$) in the two rightmost panels of each row.
 
We adopt a source-redshift distribution extending to higher redshifts than expected for LSST lens sources in order to obtain an approximately flat Einstein-radius distribution for training. As a result, the fraction of detectable host galaxies in the raw simulations is initially lower than expected ($\sim 50\%$; \citet{Ana_2024}). To mitigate this and maintain a balanced training set, we rescale the brightness of the HUDF galaxies such that approximately $50\%$ of the hosts are detectable. A host galaxy is considered detectable if its lensed surface brightness exceeds five times the background noise standard deviation and its total magnification is greater than $3$, ensuring sufficiently extended and identifiable host arcs. This enforces an approximately equal split between detectable and non-detectable hosts in the training set.

\subsection{A new negative subclass - SN Ia in the lens galaxy}
\label{sec:sn_in_lens}

We simulate an additional class of negative examples, SNe~Ia in lenses (Set~$N_2$), in which a SN~Ia occurs in the foreground lens galaxy rather than in the background source galaxy. The background galaxy, sampled from the HUDF population, is placed at redshifts drawn from the same distribution as the hosts of the mock LSNe~Ia systems (Set~$P$), and can therefore produce qualitatively similar lensed arcs. The key difference relative to the LSNe~Ia systems is that the SN~Ia itself is not lensed in Set~$N_2$, even though lensed arcs from the background galaxy may still be present.

We expect this class to yield the most challenging false positives for
the classifier, as it can be difficult to determine whether the SN
originates in the background lensed galaxy or in the foreground lens,
particularly at early times when only a single SN image might be visible in a
lensing system. We therefore explicitly include these systems during
training to allow the model to learn the subtle differences between the
two scenarios. Since the background galaxies in Set~$N_2$ follow the same
properties as the hosts in Set~$P$, approximately $50\%$ of them are
expected to be detectable. 

An interesting consequence is that, at early phases, Set~$P$ systems with detectable hosts can closely resemble Set~$N_2$ systems with detectable lensed arcs when only a single LSN image is present. Conversely, the remaining fraction of Set~$P$ systems
without detectable hosts can be more easily confused with Set~$N_3$,
corresponding to SNe~Ia occurring in LRGs. This interplay among the
negative subclasses significantly increases the classification
difficulty.

\subsection{PSF and Poisson evolution in the mock time series}
\label{sec:psf_vary}
In HOLISMOKES XVII \citep{Bag2025}, the seeing (and hence the PSF) was kept fixed across all epochs of a given mock time series and inherited from the coadded image used to seed the simulation. For simplicity, the Poisson noise realization was kept constant in time, even though the background noise realization varied between epochs. Although this assumption is computationally convenient, it introduces an artificial uniformity in the simulated time sequence. The only class for which both the PSF and Poisson noise varied across epochs was the HSC variable class, since these were not simulated but directly drawn from the HTS, i.e., from real observations. This discrepancy can introduce unintended cues that models may exploit.

In this work, we relax both assumptions and allow the seeing (or PSF) and the Poisson noise to vary across epochs when generating the mock time series. In particular, variations in the PSF and the injection of additional flux, such as SN Ia images and lensed host-galaxy arcs in lensed systems or a single SN Ia in unlensed systems, modify the corresponding Poisson noise component in the variance map. We explicitly account for this effect when constructing the mock time series, thereby providing a self consistent pathway to generate realistic time series from a static observation consisting of an image, variance map, and PSF. A brief overview of the procedure is given below, while the full details are provided in Appendix~\ref{app:flux_psf_var_map}.

\subsubsection{Sampling the seeing and PSF} \label{sec:seeing_sampling_psf}

The HTS provides a limited time-domain sampling, with
$(5, 9, 13, 13)$ observations in the $g,r,i,z$ bands, respectively, as shown in Fig. 1 of \citet{Bag2025}. To model realistic
epoch-to-epoch seeing variations in each band, we fit a Gaussian distribution to the
historical seeing values in that band and randomly draw the target seeing at each epoch
from the fitted distribution.

We model the PSF as a two-dimensional Gaussian and define its seeing via the effective full
width at half maximum (FWHM) of the fitted Gaussian with covariance matrix $\Sigma$,
\begin{equation}\label{eq:fwhm_Sig}
\mathrm{seeing} \equiv \mathrm{FWHM}_{\rm eff}
=
2\sqrt{2\ln 2}\,\left(\det \Sigma\right)^{1/4}\;.
\end{equation}
Given a target seeing, we construct the corresponding PSF via kernel convolution,
\begin{equation}\label{eq:psf_conv}
\mathrm{PSF}_{\rm target}(\vecx) = (k * \mathrm{PSF}_{\rm original})(\vecx)\;.
\end{equation}
Here, $\mathrm{PSF}_{\rm original}$ denotes the PSF of the input image prior to the seeing adjustment. The normalized Gaussian kernel $k$ is chosen such that the convolved PSF, $\mathrm{PSF}_{\rm target}$, corresponds to the sampled target seeing at that epoch. Since both the original PSF and the kernel are modeled as two-dimensional Gaussians,
$\mathrm{PSF}_{\rm target}$ is also Gaussian, with additional ellipticity and arbitrary orientation.

To capture realistic morphological variations of the PSF across epochs, we randomize the kernel orientation and ellipticity, drawing the rotation angle from $\theta_k \sim \mathcal{U}(0,\pi)$ and the axis ratio from $r_k \sim \mathcal{U}(0.8,1.0)$. This results in
moderate, time-dependent variations in PSF shape consistent with real observations. The
detailed construction of the kernel, including its covariance, ellipticity, and rotation,
is described in Appendix~\ref{app:get_kernel}.

Although the convolution-based approach restricts the target PSF to be broader than the
original one, i.e.\ $\mathrm{seeing}_{\rm target} \ge \mathrm{seeing}_{\rm original}$, it
nevertheless allows us to reproduce the HTS seeing distributions, as the ``original'' PSFs of the
coadded HSC images exhibit systematically better seeing than typical HTS observations in all
bands. Consequently, the seeing distributions across epochs in our mock time series
closely match those of the HSC variables, and the overall dataset faithfully reproduces the
observational conditions of the HTS.

\subsubsection{Implementing the PSF and Poisson noise variation at different time-steps}
Starting from a single observation characterized by the image flux
$\Ix$, the PSF, and the variance map $\var[\Ix]$ (composed of Poisson and
background components), we construct time series of lensed or unlensed
SNe, including variations in the PSF and Poisson noise. Details of this
procedure are given in Appendix~\ref{app:flux_psf_var_map}; here we
provide a brief outline.

\begin{itemize}
    \item Upon injecting additional flux (from unlensed or lensed SNe,
    or from lensed host arcs) $\Delta I(\vecx)$, we update the variance
    map as $\var[\Ix] \rightarrow \var[\Ix] + C_t\,\Delta I(\vecx)$,
    where $C_t$ is the slope of the linear fit to $\var[\Ix]$ versus
    $\Ix$, as defined in Eq.~\eqref{eq:tot_var2} in Appendix \ref{app:flux_psf_var_map}.

     \item At each time step, we sample a seeing value from the HTS
    historical distribution in each band (Sec.~\ref{sec:seeing_sampling_psf}).
    We then derive a 2D Gaussian kernel $k$ with random ellipticity and
    rotation that transforms the original PSF to the target PSF with the
    desired seeing, following Eq.~\eqref{eq:psf_conv} and the procedure
    described in Appendix~\ref{app:get_kernel}. 

    \item We then convolve $\Ix$ with the kernel, resulting in noise
    suppression in the convolved image (see Appendix
    \ref{app:flux_psf_var_map}). The target variance
    $(k*\var[I])(\vecx)$ is restored by adding uncorrelated noise drawn
    from the residual variance, following Eqs.~\eqref{eq:var_restore1}--\eqref{eq:var_restore2}:
    
    \begin{align}
        I_{\rm target}(\vecx)
        &= (k*I)(\vecx) + e_{\rm resi}(\vecx) \;, \label{eq:var_restore11} \\
        e_{\rm resi}(\vecx)
        &\sim \mathcal{N}\!\left(
        0,\,
        (k * \var[I])(\vecx) - (k^2 * \var[I])(\vecx)
        \right) , \label{eq:var_restore12}
    \end{align}
    where $k^2$ denotes the element-wise square of the kernel $k$.
    \item To ensure consistent noise variation across all time steps,
    including Poisson fluctuations, we add an additional noise layer
    drawn from the target variance map,
    $I_{\rm target}(\vecx) \rightarrow I_{\rm target}(\vecx) +
    \mathcal{N}\!\left(0,\,(k * \var[I])(\vecx)\right)$, at the expense of
    reducing the effective S/N by a factor of $\sqrt{2}$.
\end{itemize}

\subsection{Implemented SALT2-extended SN Ia lightcurves with intrinsic stretch and color}
\label{sec:satl2_imp}
In this work, we model SN Ia light curves using the empirical SALT2 model \citep{salt2}, particularly through the \texttt{salt2-extended} spectral template that provides larger wavelength coverage. This replaces the Hsiao template \citep{Hsiao2007} used in HOLISMOKES XVII and allows us to incorporate intrinsic variations in SN Ia light curves through stretch and color parameters, which affect both the luminosity and the shape of the light curves. Further details of the implementation are provided in Appendix \ref{ref:SALT2}.

Figure~\ref{fig:ts_demo} presents time series of three LSN~Ia systems as
representative examples. Each system is shown using two rows: the upper
row displays the PSF at each observing epoch, illustrating its temporal evolution following Sec.~\ref{sec:psf_vary}, while the lower row shows the corresponding image cutouts of the system. The flux evolution of the LSNe~Ia images follows the light curves shown in Fig.~\ref{fig:lc_demo}, including the effects of lensing magnification and time delays. All panels are shown in the $i$-band for demonstration. The
first two examples correspond to the systems shown in
Fig.~\ref{fig:host_light_demo} and contain clearly detectable lensed host
arcs. In contrast, the third example illustrates a configuration in
which the lensed host galaxy is not detectable at the survey depth,
leaving only the multiple SN images detectable in the time series.

Analogous time-series examples for a couple of systems belonging to the newly introduced negative subclass ($N_2$), SNe~Ia occurring in lens galaxies, are shown in Fig.~\ref{fig:sn_in_lenses_demo}. As in Fig.~\ref{fig:ts_demo}, the PSF varies across observing epochs. In the top example, the lensed background host galaxy is detectable and the arcs are clearly visible, whereas in the bottom example the background galaxy remains undetectable at the survey depth. In both cases the SN~Ia occurs in the foreground lens galaxy, illustrating how such systems can closely resemble LSNe at early phases and therefore constitute a challenging negative class for the classifier.

\subsection{Time sampling}

In addition to matching the spatial properties of the data, we ensure consistent temporal sampling across all components of the dataset. As in HOLISMOKES XVII \citep{Bag2025}, the simulated time series follow the cadence of the HTS, specifically that of the 2017 observing season. Because the cadence of the observed HSC variables (Set $N_1$) is fixed by the HTS survey schedule, this ensures that the simulated LSNe Ia and normal SNe Ia share the same cadence distribution, avoiding artificial differences in temporal sampling.


The first detection time of the simulated series is drawn uniformly from $t_0 \sim \mathcal{U}(-25,-15)$ days in the observer frame, measured relative to the $i$-band peak of the first image for lensed systems and of the SN for normal SN subclasses. For lensed systems, we additionally require that the second image appears within the constructed series. The maximum duration of a series is limited to $80$ days in any individual band (except $g$) and $120$ days when combining observations across all bands. The timestamp of the first detection is set to zero, with all subsequent observations recorded relative to this reference time to avoid revealing the true phase of the event, which would not be accessible in real observations.

\section{Model architecture and data normalization}
\label{sec:model}
We adopt the same overall deep-learning architecture, multi-band data arrangement, and normalization procedure as in HOLISMOKES XVII \citep{Bag2025}. The model consists of two parallel channels: a \texttt{ConvLSTM}-based branch that processes multi-band image time series, and an \texttt{LSTM}-based branch that ingests the corresponding observation times to learn characteristic transient time-scales. The features extracted by these two branches are concatenated and passed to a final dense layer for prediction. Owing to the increased realism and complexity of the data considered here, we re-optimized the network hyperparameters, including the number of \texttt{ConvLSTM}/\texttt{LSTM} layers and units per layer. The final model therefore employs a larger number of recurrent cells to adequately capture the enhanced data complexity and achieve optimal performance.

Without loss of generality, each sample is constructed to contain exactly 14 observations distributed across the $griz$ bands, enabling a fair comparison of the classification results after each observing epoch. Following the notation introduced in HOLISMOKES XVII, we denote the cumulative number of observations by $\Ne$, with $\Ne \in {1,2,\dots,14}$, representing the progressive accumulation of multi-band observations up to a given point in time. Asynchronous observations are arranged into inputs of fixed shape, with missing bands zero-padded to enable uniform multichannel processing. Images are normalized relative to the first observation to preserve color information while maintaining numerical stability for real-time inference. Since the preprocessing pipeline remains unchanged, we refer the reader to \citet{Bag2025} for full implementation details.


\section{Results}
\label{sec:results}

\begin{figure*}
    \centering
    \includegraphics[width=\textwidth]{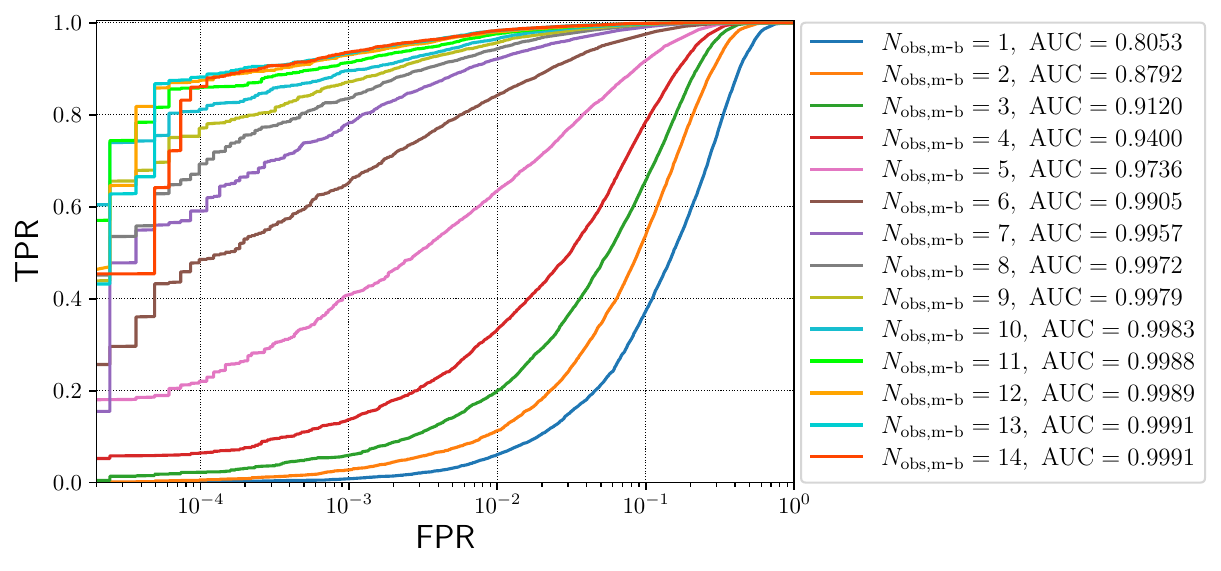}
    \caption{Receiver operating characteristic (ROC) curves for the multi-band classification results obtained from $14$ observations per sample. The corresponding area under the curve (AUC), which approaches unity for a perfect classification, is indicated in the legend. Note that the ROC curve improves rapidly over the first few observations, suggesting increased model 
    accuracy with early observations, and begins to saturate later in the series. }
    \label{fig:ROC0}
\end{figure*}

\begin{figure*}
    \centering
    \includegraphics[width=\textwidth]{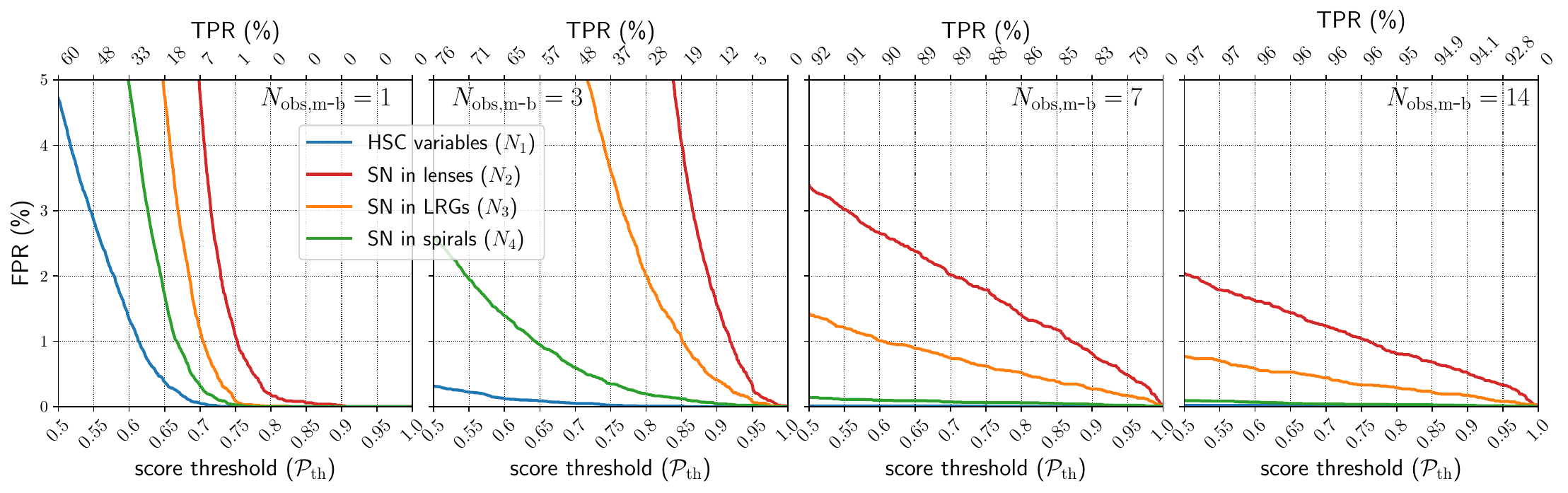}
    \caption{Each panel shows FPR as a function of score threshold separately for different negative components. 
    The four panels represent the results obtained after 1st, 3rd, 7th, and 14th observation, respectively. TPR values corresponding to different score thresholds are marked on the top $x$-axis. The FPR for all four negative components -- HSC variables, SN exploding in lenses, LRGs, and spiral galaxies -- gradually decreases with additional observations. However, it is clearly evident that SN in lenses and LRGs dominate the FPR budget, as they are more likely to be confused with the LSNe.}
    \label{fig:neg_fprs}
\end{figure*}

\begin{figure*}
    \centering
    \includegraphics[width=\textwidth]{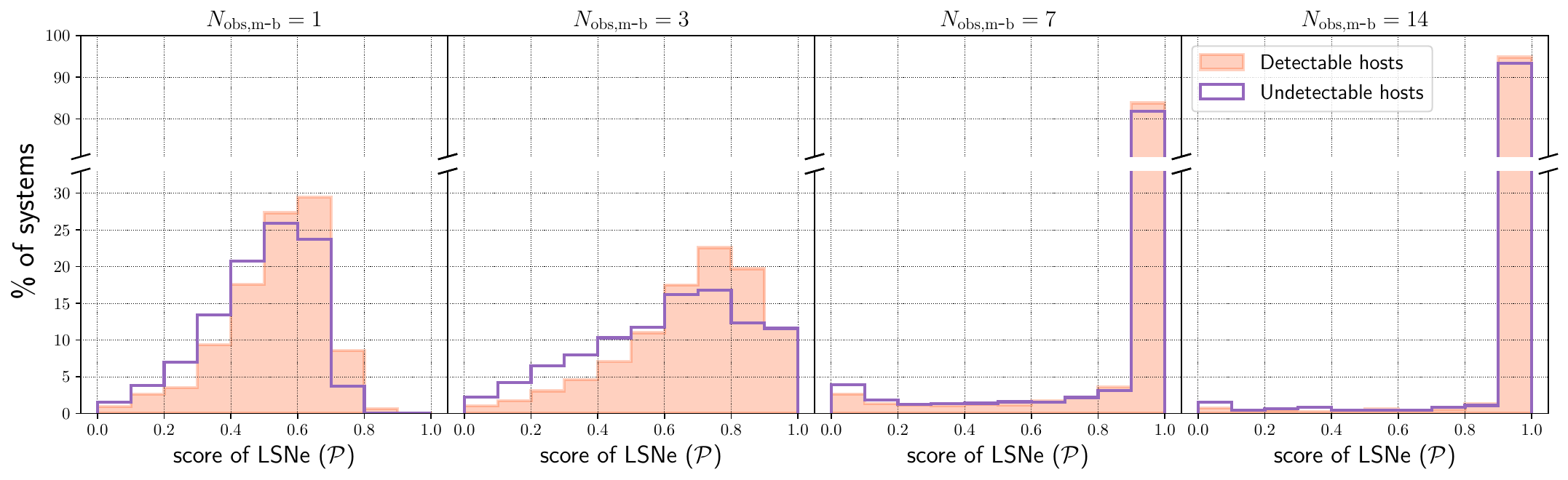}
    \caption{Comparison of model-predicted score distributions within the positive LSNe Ia class, split into systems with detectable and undetectable host galaxies. The four panels show the results after the 1st, 3rd, 7th, and 14th observing epochs ($\Ne$).}
    \label{fig:bright_vs_dim_host_arcs}
\end{figure*}

\begin{figure*}
    \centering
    \includegraphics[width=\textwidth]{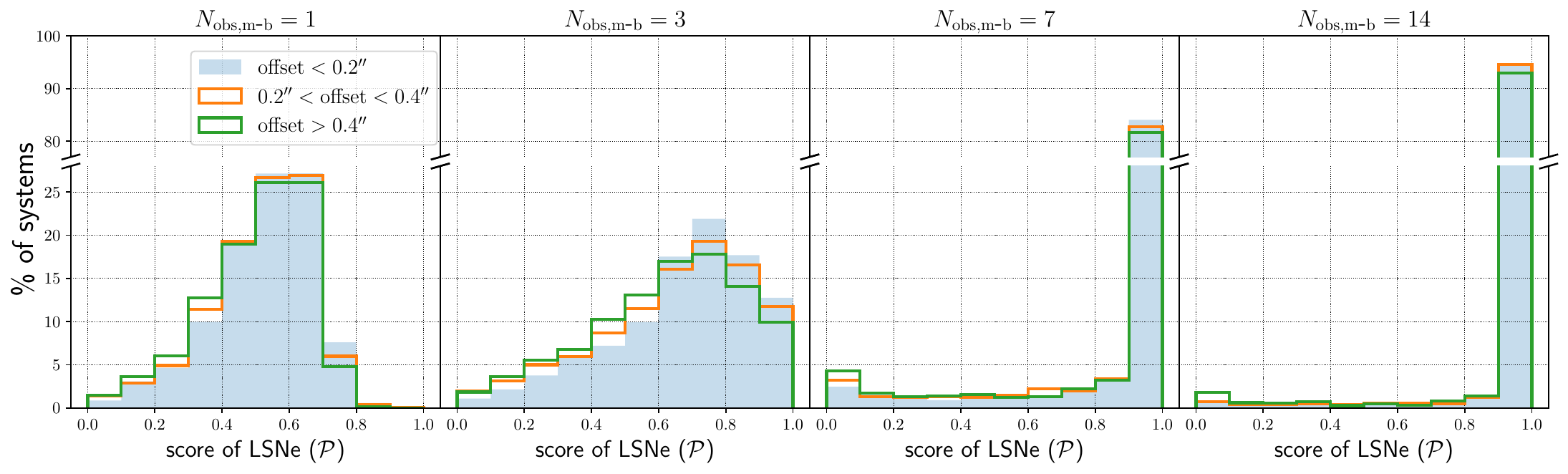}
    \caption{Distribution of model-predicted scores for LSNe Ia divided into three bins of offset between the SN and the host galaxy centre. The four panels show the results after the 1st, 3rd, 7th, and 14th observing epochs ($\Ne$).}
    \label{fig:score_offset}
\end{figure*}
\begin{figure}
    \centering
    \includegraphics[width=\linewidth]{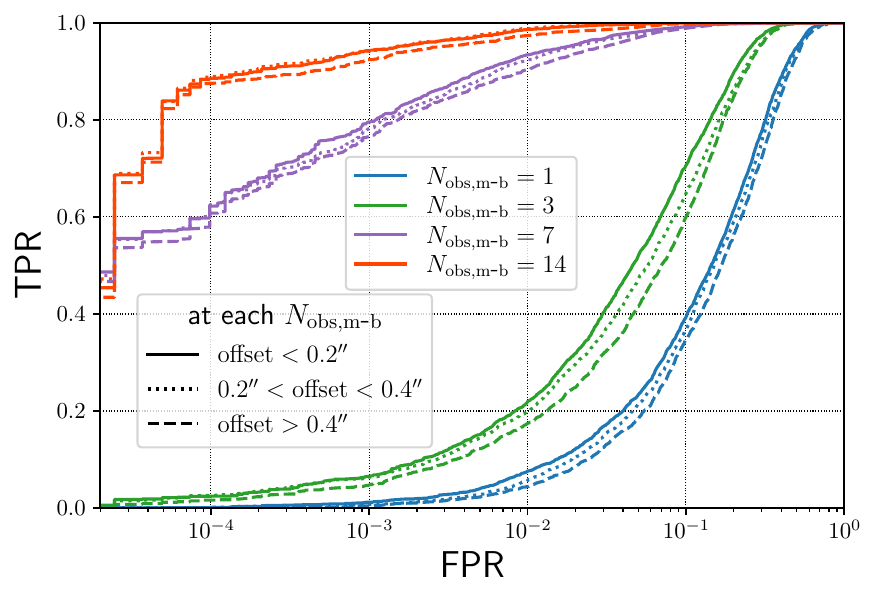}
    \caption{ROC curves comparing classification performance for LSNe Ia in different SN-host offset bins after different observing epochs ($\Ne$).}
    \label{fig:ROC_offset}
\end{figure}

\subsection{Overall network performance}
The classification performance is illustrated in Fig.~\ref{fig:ROC0} using receiver operating characteristic (ROC) curves, which show the true positive rate (TPR) as a function of the false positive rate (FPR) while varying the decision threshold on the predicted score. Fig.~\ref{fig:ROC0} presents 14 ROC curves corresponding to predictions made after each of the 14 observations, i.e., at $\Ne = 1,2,\ldots,14$, for all test samples. The corresponding area under the curve (AUC), which approaches unity for a perfect classifier, is indicated in the legend.

The ROC curves improve rapidly during the first few observations, indicating that the model quickly becomes effective at distinguishing the lensed systems from other transients as early data accumulate. This improvement is also reflected in the corresponding AUC values. By the seventh epoch, the model reaches a TPR of $\sim 60\%$ at an FPR of $\mathcal{O}(10^{-4})$, increasing to $\sim 80\%$ by the tenth epoch. Beyond this point the ROC curves begin to saturate, with only marginal improvement from additional observations.

These results demonstrate that the model achieves strong performance with relatively early observations, enabling timely identification of candidates for follow-up. Importantly, the overall performance remains comparable to that reported in HOLISMOKES XVII \citep{Bag2025}, despite the increased realism of the present simulations, including time-varying PSF conditions, the inclusion of Poisson noise, and the addition of a challenging negative class corresponding to SN in lenses (Set $N_2$).

Our negative samples consist of several subclasses with varying levels of difficulty to distinguish from the positive lensed cases. These include HSC variables ($N_1$), SN Ia in lenses ($N_2$), SN Ia in LRGs ($N_3$), and SN Ia in spiral galaxies ($N_4$), as summarized in Table~\ref{tab:details}. Since these subclasses exhibit different observational characteristics, it is instructive to examine which of them contribute most to the FPR and to what extent.

Figure~\ref{fig:neg_fprs} shows the FPR as a function of the score threshold ($\mathcal{P}_{\rm th}$) for each negative subclass. The four panels correspond to the model performance after $\Ne = 1, 3, 7$, and $14$. In each panel, the curves represent the cumulative score distributions of the four negative subclasses, directly indicating their contribution to the total FPR at a given threshold. The top $x$-axis shows the corresponding TPR for the same threshold values.

SN~Ia in lenses ($N_2$) and in LRGs ($N_3$) dominate the FPR budget, reinforcing our choice of oversampling them in the training set. These configurations are intrinsically more difficult to distinguish from genuine lensed systems because they occur in massive galaxies that resemble typical lens environments. Even at later epochs they remain the primary contributors to the FPR, while HSC variables ($N_1$) and SN~Ia in spirals ($N_4$) contribute only marginally.

Initially, between $\Ne=1$ and $\Ne=3$, the FPR for $N_2$ and $N_3$ increases over part of the threshold range, although the accessible TPR range also expands substantially over the same interval. With only a few observations the model lacks sufficient temporal information, causing a larger fraction of these systems to receive high scores similar to genuine lensed events. As additional observations become available, the classifier better separates these cases and the corresponding FPR decreases, as seen when comparing the panels for $\Ne=7$ and $\Ne=14$. Overall, the FPR for all subclasses decreases with $\Ne$ as the model processes more observations.

At higher score thresholds, even the contamination from the FPR-dominating subclasses becomes small. For $\mathcal{P} > 0.8$, the FPR from $N_2$ falls below $\sim1.5\%$ at $\Ne=7$ and $\sim0.8\%$ at $\Ne=14$, while for $N_3$ it drops below $\sim0.5\%$ and $\sim0.2\%$ at the same epochs.
In realistic survey populations these subclasses are not necessarily rarer than LSNe themselves; in particular, $N_3$ is expected to be several orders of magnitude more abundant than LSNe~Ia prior to any downstream filtering and could therefore represent an important source of false positives.\footnote{The FPR measures the fraction of objects within a given negative subclass that are misclassified as positive. The ``false positive contribution'' accounts for the abundance of that subclass in the population and represents its expected contribution to the total number of false positives.} The low FPR achieved for these subclasses therefore underscores the importance of suppressing such contaminants and demonstrates that the classifier effectively controls the most challenging sources of false positives while maintaining strong detection performance.

\subsection{Classification performance: Detectable vs. Undetectable LSNe host galaxies}
Next, we investigate how the presence of lensed host-galaxy arcs affects the classification performance. For this purpose, we divide the positive samples into two subsets: systems with detectable host arcs and systems where the host remains undetectable at the survey depth. Figure~\ref{fig:bright_vs_dim_host_arcs} shows the score distributions for these two subsets, compared with the full sample (shown as the filled histogram), at $\Ne = 1, 3, 7$, and $14$ in the four panels.

At early epochs, LSNe Ia with detectable host arcs tend to receive higher scores, as the arcs provide additional spatial features that aid identification of the lensing configuration. However, this advantage gradually diminishes as more observations accumulate and the model gains information from the evolving time series. By later epochs, both subsets exhibit nearly identical score distributions, with only a slight advantage for LSNe Ia with detectable hosts.

The two subsets are also susceptible to different types of confusion with the negative classes, particularly during the early stages of the observation sequence. Lensed systems with detectable host arcs are primarily confused with the subset of the ``SN in lenses'' negative subclass ($N_2$) that also exhibits detectable background lensed galaxy arcs. In contrast, lensed systems without detectable hosts more closely resemble the remaining $N_2$ population as well as the ``SN in LRGs'' subclass ($N_3$). Despite these different sources of ambiguity, the model achieves comparable performance for both subsets by the end of the time series.

\subsection{Performance at different offsets between the LSNe and their host galaxy}

The offset between the SN and the centre of its host galaxy can influence whether the host galaxy is also lensed and detectable, potentially affecting classification performance. In our simulations, the SN is placed at a distance drawn from $\sim \mathcal{U}(0,3R_{50})$, clipped at $2''$, exposing the model to LSNe across a wide range of SN-host offsets.

In this section we investigate how the trained model performs for different SN-host offsets. The test sample is divided into three offset bins: ${\rm offset}<0.2''$, $0.2''<{\rm offset}<0.4''$, and ${\rm offset}>0.4''$, with each subset containing approximately $2$--$3$ thousand systems. Figure~\ref{fig:score_offset} shows the score distributions for these subsets, with the four panels corresponding to results after the 1st, 3rd, 7th, and 14th observations. Overall, we do not observe any significant difference in the score distributions across the offset bins at any epoch. However, systems with smaller offsets, particularly ${\rm offset}<0.4''$, tend to receive slightly higher scores. This is likely due to the stronger spatial association between the SN and the lensed host-galaxy light, which increases the spatial correlation in the images.

This trend becomes clearer in Fig.~\ref{fig:ROC_offset}, which shows the ROC curves for the different offset subsets. Each colour corresponds to a different observing epoch, while different line styles represent the offset bins. Consistently, systems with smaller offsets show slightly better ROC performance, although the difference remains small. This behaviour is consistent with our earlier result that the overall classification performance is only weakly affected by the detectability of the host galaxy (see Fig.~\ref{fig:bright_vs_dim_host_arcs}). Overall, the model performs robustly across a wide range of SN-host offsets. 

\subsection{Additional trends in line with HOLISMOKES XVII}
We also find that quad systems are generally easier to identify than doubles. Lensed systems with larger image separations are likewise detected more reliably, since the multiple images are more clearly resolved. A subtle caveat arises for some widely separated double systems: larger separations typically correspond to longer time delays, so the second-arriving image may appear much later in the time series. Consequently, classification can be more challenging at early epochs, before the arrival of the second image. 

In this context, LSNe that receive low scores ($\mathcal{P}<0.1$), often doubles, typically have one or more images blended with the lens light or fainter image(s) missed due to sparse time sampling from cadence limitations. Such configurations can resemble SNe in lenses ($N_2$) or in LRGs ($N_3$), leading to missed detections and, conversely, allowing some $N_2$ and $N_3$ systems to be assigned high scores.

As expected, the multi-band model, which processes observations from any filter, outperforms the single-band models that rely on observations from only one band. This improvement arises from the additional color information and the richer temporal context captured by the multi-band model. These trends are consistent with the findings of HOLISMOKES XVII \citep{Bag2025}, and are therefore not shown explicitly here.

\subsection{Sibling supernovae as a source of confusion}
\label{sec:sibling_sne_main}
Finally, we investigate a potential source of confusion for LSNe searches: sibling SNe, i.e. multiple SNe occurring in the same galaxy. Sibling SNe can provide opportunities for studies of SN progenitor diversity, tests of SN Ia standardization and intrinsic scatter, and investigations of host-galaxy environmental effects, since the events occur in the same galaxy and share the same distance and global environment \citep{brown2014, Scolnic2020, Kelsey2023}. If two such events occur within a sufficiently short time interval that both are visible simultaneously, e.g. \citep{Stritzinger2002}, they may visually resemble multiply imaged transients. We therefore evaluate the model response on a dedicated sample of simultaneously visible sibling SNe that was not included during training, as these events are themselves rare and scientifically interesting transients and their occasional selection as candidates would be acceptable. While a small fraction of such sibling SNe receive high lensing scores, primarily for configurations where the two SNe lie at large projected angles with respect to the galaxy centre ($\theta \gtrsim 90^\circ$), the majority are correctly classified as unlensed. The construction of this test sample and the detailed classification results are presented in Appendix \ref{sec:sibling}.

\section{Conclusions and discussion}
\label{sec:conclusions}

In this work, we extend the image-time-series framework introduced in HOLISMOKES XVII \citep{Bag2025} to detect lensed supernovae (LSNe) directly from survey alert streams. The model operates directly on multi-band time series of image cutouts using a \texttt{ConvLSTM}-based architecture, enabling it to capture spatial and temporal correlations simultaneously. The primary goal is identification of LSNe early in their evolution from real-time survey alerts.

The present work improves upon HOLISMOKES XVII by relaxing several simplifying assumptions and introducing additional astrophysical and observational realism into the simulations and training datasets. We update the dataset from HSC PDR2 to PDR3 observations, resulting in an $\sim 50\%$ increase in the available input data. Lensed host-galaxy arcs are incorporated into the simulated LSNe, extending the model to cases where lensed host arcs are detectable in the imaging data. We also introduce epoch-to-epoch variations in the PSF and include Poisson noise consistent with changing observing conditions. Taken together, these improvements enable a self-consistent framework for generating realistic time-series observations from a single static observation consisting of an image, variance map, and PSF.
The transient modeling is improved using SALT2-based SN~Ia light curves with intrinsic stretch and color variations. Finally, we introduce an additional negative class consisting of SNe occurring in the foreground lens galaxy itself, which represents a realistic and challenging source of false positives in LSNe searches.

The ROC curves show a rapid improvement in classification during the first few observations, indicating that the model quickly accumulates sufficient information to distinguish lensed from unlensed events. By the seventh observation ($\Ne=7$), the model achieves a TPR of $\sim60\%$ at an FPR of $\mathcal{O}(10^{-4})$, increasing to $\gtrsim80\%$ by the tenth observation ($\Ne=10$). Beyond this point the performance begins to saturate, suggesting that most of the relevant information is already captured during the early part of the time series. Notably, these results remain broadly comparable to those obtained in HOLISMOKES XVII despite the additional observational and astrophysical complexities introduced in this work, demonstrating that the method remains effective for early identification of LSNe.

A closer inspection of the predictions for the different negative subclasses shows that SNe in lens galaxies and in LRGs yield the highest FPRs. These systems can resemble lensed events under certain configurations, particularly for doubly imaged systems where the multiplicity is not clearly resolved or only a single image is visible at early epochs. Including ``SN~Ia in lenses'' as an explicit negative class is astrophysically motivated and improves the model’s robustness to the most challenging contaminants. In practice, the model effectively controls the FPR from these subclasses at conservative score thresholds, e.g., ($\mathcal{P}_{\rm th} > 0.8$).

We also examined the impact of detectable host-galaxy arcs in LSNe (positive samples) and found that they provide additional spatial features that improve classification at early epochs. As more observations accumulate, however, this advantage diminishes and systems with detectable and undetectable hosts achieve comparable classification scores. We further investigated how the classification performance depends on the offset between the LSN and the centre of its host galaxy. Overall, the model performs robustly across a wide range of offsets, with only a mild dependence on SN-host separation. Systems with smaller offsets are classified slightly more efficiently, likely because the stronger spatial association between the SN and the lensed host galaxy light enhances the spatio-temporal correlations in the image sequence. Nevertheless, the differences remain modest, indicating that the method is largely insensitive to the exact SN location within the host galaxy.

We further investigated potential confusion with sibling SNe, particularly when two independent SNe occur in the same galaxy within a comparable time window. Although a fraction of such systems can mimic lens-like configurations and receive high classification scores, the majority are still correctly identified as unlensed events. Even without including sibling SNe during training, the model appears able to exploit additional information such as colour evolution, brightness ratios, and consistency with realistic lensing configurations. These results suggest that the classifier can generally distinguish such systems from genuine LSNe, although a small fraction may still be flagged as candidates.

We also explored how different lens configurations affect classification performance. Consistent with HOLISMOKES XVII, quad systems are generally easier to identify than doubles, and systems with larger image separations are detected more reliably. However, some widely separated doubles exhibit large time delays, making early identification more difficult before the second image appears.

Overall, these results demonstrate that the image-time-series approach introduced in HOLISMOKES XVII remains robust when extended to more realistic observational conditions and astrophysical configurations. The ability to maintain strong early-detection performance in the presence of varying PSF conditions, Poisson noise, host-galaxy arcs, and additional negative classes highlights the potential of this method for real-time LSNe searches in time-domain surveys. In particular, the framework is well suited for application to the LSST alert stream, where the improved cadence relative to the HTS-like cadence adopted in this work is expected to further improve the early identification of LSNe and enable timely follow-up observations.

In the present work we have used HSC observations as a proxy for LSST and explored most of the realistic effects that can be modeled within this framework. The next step will be to transition to LSST data products and incorporate the specific characteristics of the LSST alert stream, including survey-specific observational nuances arising from its massive alert volume. In addition, the present study focuses on Type Ia LSNe, but the framework can be naturally extended to other types, e.g. core-collapse SNe, and potentially to other transients such as tidal disruption events (TDEs) or kilonovae (KNe) by incorporating the corresponding light-curve models. A key challenge will be capturing realistic light-curve diversity while maintaining adequate coverage of lensing configurations across a broad range of source redshifts. These developments will form the focus of the next stage of this program.

\begin{acknowledgements} 
We thank Christian Vogl, Justin Pierel, Tim Meinhardt and 
Laura Leal-Taixe for many useful discussions. This work is funded by the Deutsche Forschungsgemeinschaft (DFG, German Research Foundation) - SFB 1258 - 283604770. SB also acknowledges the support provided by the Alexander von Humboldt Foundation.
SHS thanks the Max Planck Society for support through the Max Planck Fellowship. 
This work is supported in part by the Deutsche Forschungsgemeinschaft (DFG, German Research Foundation) under Germany's Excellence Strategy -- EXC-2094 -- 390783311. RC acknowledges support from the French government under the France 2030 investment plan, as part of the Initiative d’Excellence d’Aix-Marseille Université - A*MIDEX (AMX-23-CEI-088). 
\end{acknowledgements} 

\vspace*{-6 mm}
\bibliographystyle{aa}
\bibliography{ref1}
\appendix

\section{Details of Observational Inputs}
\label{app:ingredient_details}
\subsection{Selection of LRGs from HSC wide layer}
\label{app:LRG_selection}
In this work, the parent LRG sample is drawn from the HSC PDR3 Wide layer (see \citet{Aihara2022}), with spectroscopic redshifts and velocity dispersion measurements from SDSS-IV DR14 \citep{abolfathi18, bautista18}. Applying a selection of $i$-band Kron radii $>0.8,\arcsec$, excluding galaxies with non-smooth morphologies based on the classifications of \citet{walmsley22} and \citet{walmsley23}, and imposing additional quality cuts using HSC flags and artifact rejection yields a final sample of 119,378 LRG cutouts. This is a factor $>3$ larger than the sample used in HOLISMOKES XVII \citep{Bag2025}, which was based on the smaller HSC Wide PDR2 footprint. A subset of these LRGs is used to construct 113,000 lens-source pairs, following the OM10 \citep{om10, Oguri2018} redshift distributions and enforcing a flat Einstein-radius distribution in the range $0.1$--$2.0,\arcsec$.

\subsection{HSC Transient Survey (HTS)}
\label{app:HTS}
The HSC Transient Survey (HTS) \citep{yasuda19} is a cadenced Subaru program targeting ultra-deep fields; here we focus on the COSMOS tract observed between November 2016 and April 2017. To better match LSST single-epoch depth, we constructed a shallower dataset by stacking a subset of exposures per epoch using pipeline-reduced images (see \citet{Bag2025} for more details).
\citet{chao21} compiled over $100,000$ variable sources from the HTS, which we include as negative samples, selecting only those with sufficient multi-epoch coverage from the 2017 observing window. After quality cuts, $\sim 9,000$ variables are retained and expanded via eight-fold data augmentation.

\subsection{Matching LRG properties across dataset subclasses}
\begin{figure*}
    \centering
    \includegraphics[width=\textwidth]{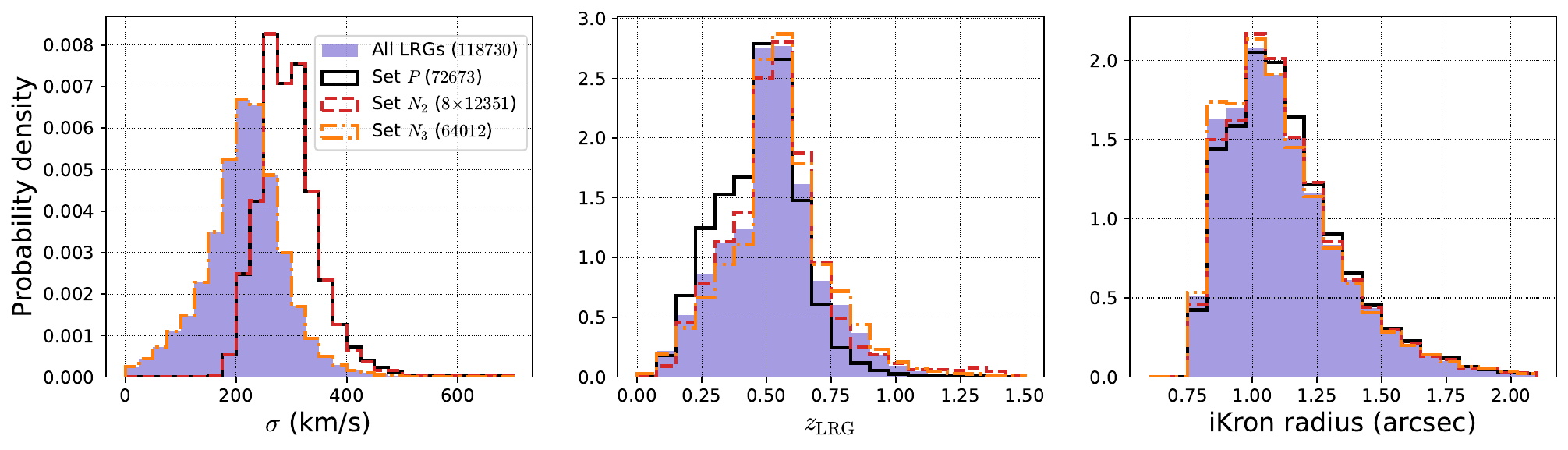}
    \caption{Distributions of velocity dispersion ($\sigma$), redshift, and
\texttt{iKron} radius (from left to right) for LRGs belonging to the
different classes considered in this analysis. For the negative subclass $N_2$, the original LRG sample was augmented eightfold to obtain a sufficient number of systems.}
    \label{fig:lrg_dists}
\end{figure*}

We collect LRGs from the HSC wide layer and ingest them into several 
subclasses of our dataset, in particular LSNe~Ia systems ($P$), SNe~Ia 
occurring in lens galaxies ($N_2$), and SNe~Ia in LRGs ($N_3$). Since these 
subclasses are constructed from the same parent LRG sample, it is important 
to ensure that their properties are matched appropriately. We therefore 
assign LRGs to the different subclasses in a controlled manner, avoiding 
duplication across components while preserving the desired property 
distributions and keeping the subclasses as large as possible.

The resulting distributions of velocity dispersion ($\sigma$), redshift 
($z_{\rm LRG}$), and iKron radius are shown in Fig.~\ref{fig:lrg_dists}. Because lens 
galaxies preferentially correspond to systems with higher velocity 
dispersions, the $\sigma$ distributions of the $P$ and $N_2$ samples are 
constructed to match each other. In contrast, for the $N_3$ subclass we 
require the velocity dispersion distribution to follow that of the parent 
LRG population in the HSC wide layer, reflecting the assumption that any 
LRG may host a normal SN~Ia. The redshift and iKron radius distributions are 
already broadly consistent across the subclasses and therefore require no 
additional selection, as shown in Fig.~\ref{fig:lrg_dists}.

\section{Varying PSF shape and orientation}\label{app:get_kernel}
Here we describe the construction of the convolution kernel $k$ used to obtain the PSF corresponding to a target seeing value through Eq.~\eqref{eq:psf_conv}. Let the original PSF be characterized by its second-moment (covariance) matrix
$\Sigma_{\rm original}$, obtained from a two-dimensional Gaussian fit. For any kernel $k$, the covariance matrix of the convolved PSF satisfies
\begin{equation}
\Sigma_{\rm target} = \Sigma_{\rm original} + \Sigma_k \;,
\end{equation}
where $\Sigma_k$ denotes the covariance matrix of the kernel. This relation holds
generally at the level of second moments, independent of the detailed functional form of the PSF.

In order to relate the covariance matrix to a scalar seeing value, we adopt the Gaussian approximation for the PSF and define the effective seeing via the determinant of the covariance matrix as given in Eq.~\eqref{eq:fwhm_Sig}. Under this approximation, imposing a desired
target seeing leads to 
\begin{equation}\label{eq:fwhm_Sig_target}
\det\!\left(\Sigma_{\rm original} + \Sigma_k\right)
=
\left(\frac{\mathrm{seeing}_{\rm target}}{2\sqrt{2\ln 2}}\right)^4 .
\end{equation}
When the original PSF is well approximated by a 2D Gaussian, this construction ensures that the resulting PSF matches the target seeing as defined by the effective FWHM in Eq. \eqref{eq:fwhm_Sig}; for a non-Gaussian PSF$_{\rm original}$, the procedure enforces the target seeing at the level of second moment.

For simplicity, we model both the original PSF and the kernel as two-dimensional Gaussians,
such that the target PSF can also be approximated by a Gaussian. Within this choice, we allow the kernel to
introduce additional ellipticity and arbitrary orientation in
$\mathrm{PSF}_{\rm target}$.
We parameterize the kernel covariance matrix as an elliptical Gaussian,
\begin{equation}
\Sigma_k
=
\sigma_{x,k}^2\,
R(\theta_k)
\begin{pmatrix}
1 & 0 \\
0 & r^2
\end{pmatrix}
R^{\mathsf T}(\theta_k),
\end{equation}
where $r=\sigma_{y,k}/\sigma_{x,k}$ is the axis ratio of the kernel and $\theta_k$ its
orientation. The rotation matrix is given by
\begin{equation}
R(\theta_k)
=
\begin{pmatrix}
\cos\theta_k & -\sin\theta_k \\
\sin\theta_k & \cos\theta_k
\end{pmatrix}.
\end{equation}

For fixed values of $(r, \theta_k)$, Eq. \eqref{eq:fwhm_Sig_target} can be written as a quadratic equation in $\sigma_{x,k}^2$. Writing
$\Sigma_k = \sigma_{x,k}^2 M$, where
$M = R(\theta_k)\,\mathrm{diag}(1,r^2)\,R^{\mathsf T}(\theta_k)$, the determinant in Eq. \eqref{eq:fwhm_Sig_target} expands as
\begin{align}
\det(\Sigma_{\rm original})
&+
\sigma_{x,k}^2\,\mathrm{Tr}\!\left[\mathrm{adj}(\Sigma_{\rm original})\,M\right]
+
\sigma_{x,k}^4\,\det(M) \nonumber\\
&=
\left(\frac{\mathrm{seeing}_{\rm target}}{2\sqrt{2\ln 2}}\right)^4 .
\end{align}
This equation admits a unique positive solution for $\sigma_{x,k}^2$ provided
$\mathrm{seeing}_{\rm target} \ge \mathrm{seeing}_{\rm original}$, ensuring that the kernel corresponds to a physical PSF broadening. The kernel widths are then given by
$\sigma_{x,k}$ and $\sigma_{y,k}=r\,\sigma_{x,k}$, which uniquely determine the kernel
covariance matrix $\Sigma_k$. Finally, the kernel is normalized such that $\sum_{\vecx} k(\vecx) = 1$, thereby preserving total flux under convolution.

\section{How injecting sources and variation of PSF affects the Poisson noise}\label{app:flux_psf_var_map}

We consider an observation of a field, described by the observed image $\Ix$, the point spread function (PSF), and the corresponding variance map $\var[\Ix]$, where $\Ix$ is the measured flux at pixel position $\vecx$ and $\var[\Ix]$ its variance. The total variance can be decomposed as
\begin{equation}\label{eq:tot_var}
\var[\Ix]=\varp[\Ix]+\varo[\Ix]\;,
\end{equation}
where $\varp[\Ix]$ denotes the Poisson variance arising from photon counting statistics and
therefore scales with the flux, while $\varo[\Ix]$ includes all additional flux-independent
contributions, such as sky background, detector noise, and other instrumental systematics,
which we collectively refer to as `background noise'.

In this section, we examine how the variance map is modified by the injection of an additional
source at a given location and by changes in the effective PSF of the image.

\subsection{Modeling Poisson noise from the observation --  $I(\vecx)$ and $\var[\Ix]$}\label{sec:psf_Poisson_noise}

\begin{figure*}
    \centering
    \includegraphics[width=0.49\textwidth]{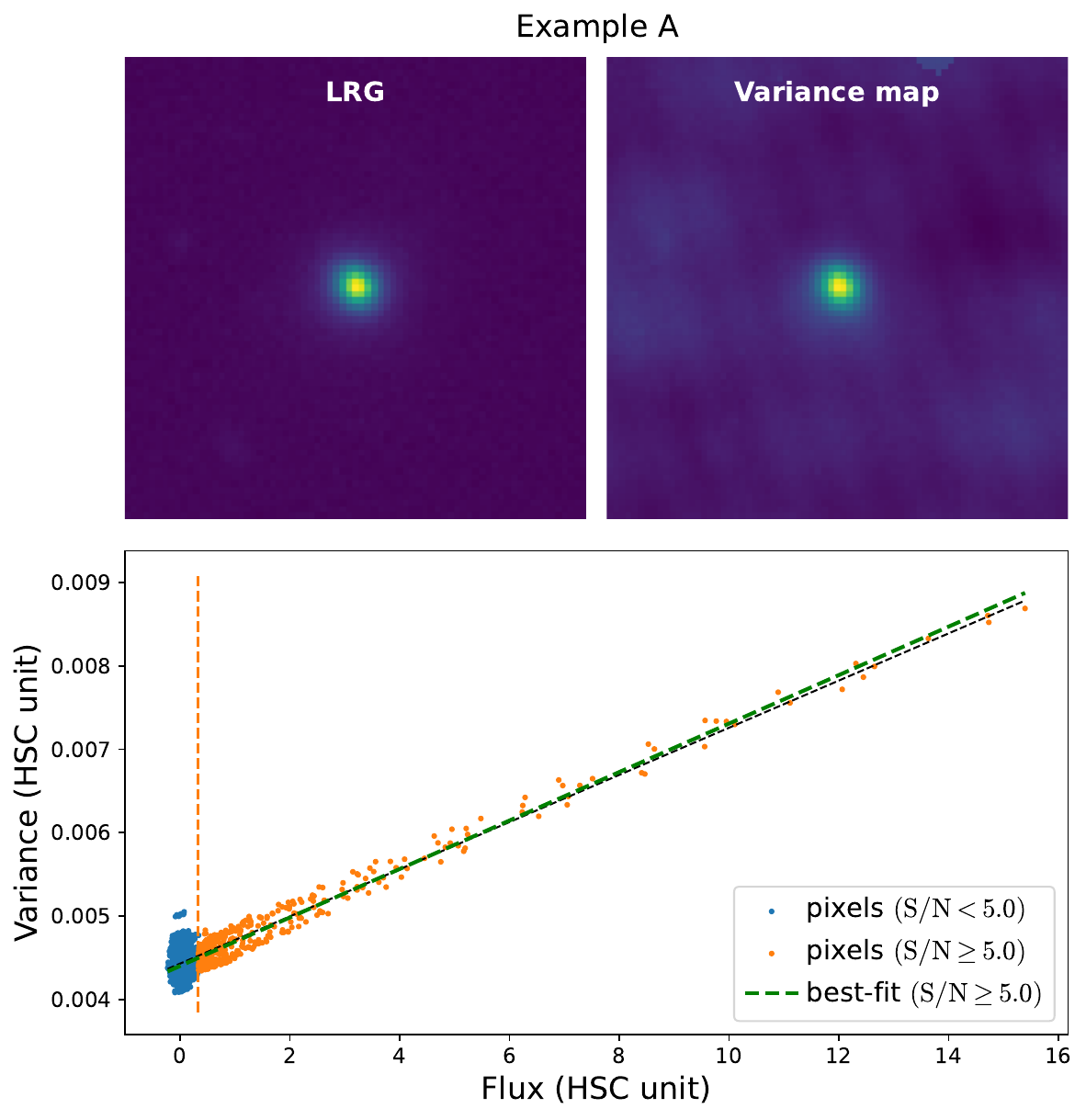}
    \includegraphics[width=0.49\textwidth]{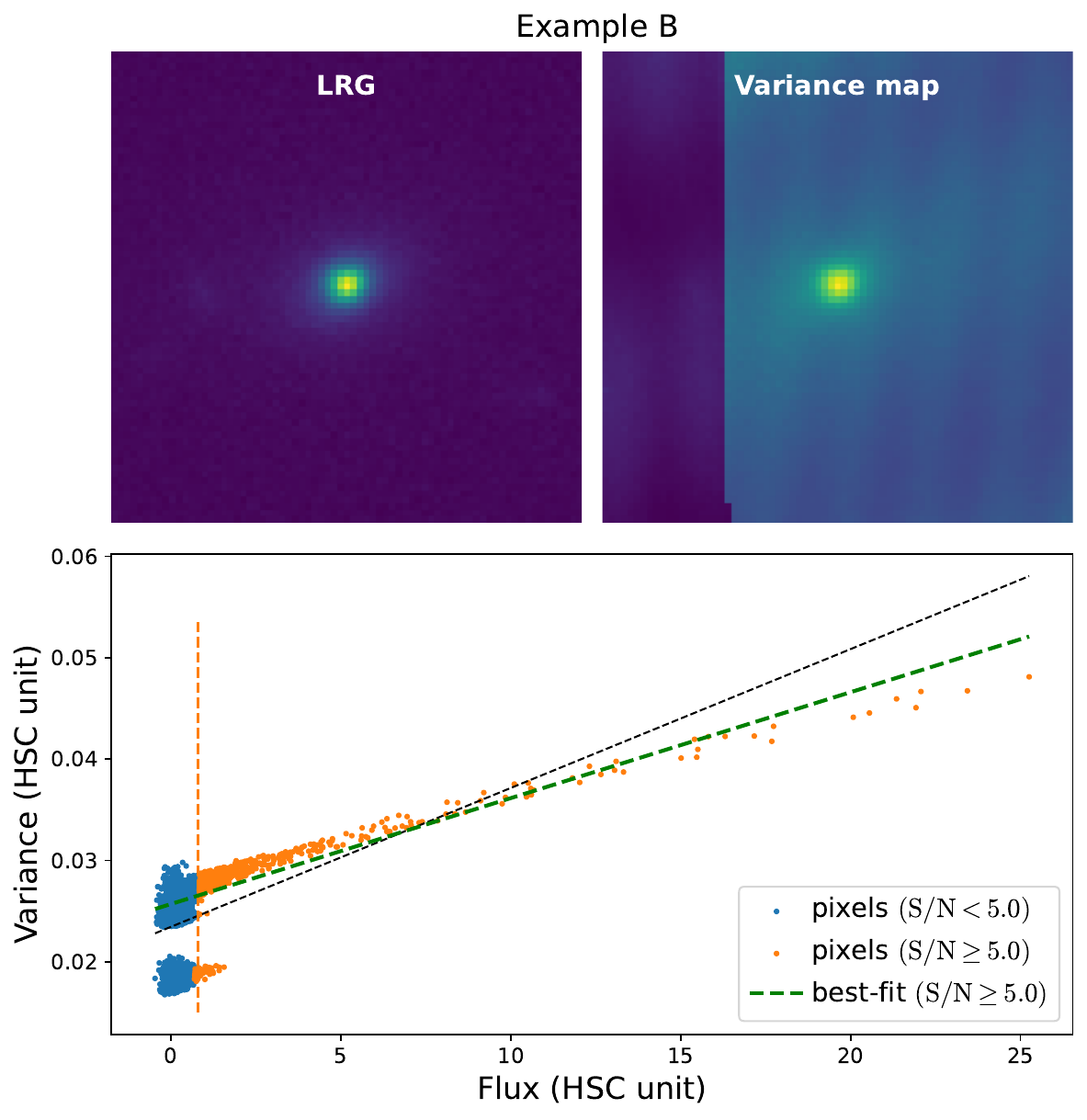}
\caption{
Pixel-level variance as a function of flux for two representative examples (LRG cutouts are shown for illustration, although the method is generic). 
For each example, the top panels display the flux image $I(\vecx)$ (left) and the corresponding variance map $\mathrm{Var}[I(\vecx)]$ (right), while the bottom panel shows the pixel-wise variance-flux relation. 
Blue and orange points indicate pixels with ${\rm S/N}<5$ and ${\rm S/N}\geq5$, respectively. The black dashed line represents a linear fit to all pixels, while the green dashed line shows the best-fit relation using only high-S/N pixels. The vertical orange dashed line marks the flux corresponding to ${\rm S/N}=5$. A clear separation of the high-S/N pixels beyond this threshold indicates a consistent linear relation.
In the left example (corresponding to the first example in Figs.~\ref{fig:host_light_demo} and \ref{fig:ts_demo}), the variance scales linearly with flux across the full dynamic range, and a single linear model provides an excellent fit. 
In the right example (corresponding to the third example in Fig.\ref{fig:ts_demo}), the variance map exhibits spatially patchy background structure, likely associated with readout or processing artifacts in HSC PDR3, producing multiple approximately linear sequences in the variance-flux plane with similar slopes but different effective background levels. Nevertheless, the high-S/N pixels remain well described by a linear relation. Since the variance correction depends only on the slope of this relation, the Poisson scaling can still be robustly determined.
}
    \label{fig:var_fit_demo}
\end{figure*}

Suppose the flux at pixel position $\vecx$ is estimated from the total photon count $N_t(\vecx)$
accumulated over an exposure time $t$ as
\begin{equation}
\Ix = C\,\frac{N_t(\vecx)}{t} \;,
\implies
\langle \Ix \rangle = \left(\frac{C}{t}\right)\langle N_t(\vecx) \rangle ,
\end{equation}
where $\langle N_t(\vecx) \rangle$ is the expected photon count in time $t$ and $C$ is the calibration constant converting photon counts to flux units, incorporating the detector gain and other instrumental calibration factors. When Poisson noise dominates, the photon count follows
\begin{equation}
N_t(\vecx) \sim \mathrm{Poisson}\!\left(\langle N_t(\vecx) \rangle\right),
\end{equation}
which implies
\begin{equation}
\var[N_t(\vecx)] = \langle N_t(\vecx) \rangle
= \left(\frac{t}{C}\right)\langle \Ix \rangle .
\end{equation}
The resulting Poisson variance of the flux is therefore
\begin{equation}\label{eq:varp_Ix}
\varp[\Ix]
= \left(\frac{C}{t}\right)^2 \var[N_t(\vecx)]
= \left(\frac{C}{t}\right)\langle \Ix \rangle \;.
\end{equation}
Equation~\eqref{eq:tot_var} can therefore be written as
\begin{equation}\label{eq:tot_var2}
\var[\Ix]
= \underbrace{\left(C/t\right)}_{\text{slope}\equiv C_t}\,\langle \Ix \rangle
+ \underbrace{\varo[\Ix]}_{\text{intercept}}\;.
\end{equation}
Assuming that the background noise is spatially uniform, the Poisson and background
contributions can be separated by fitting a straight line to the variance map $\var[\Ix]$ as a function of
the observed flux $\Ix$ (an unbiased estimator of $\langle \Ix \rangle$) over all
pixels. The slope, say $C_t$, quantifies the linear scaling of the Poisson variance with the
expected flux and the intercept yields the background variance. Consequently, when an
excess flux $\Delta \Ix$ is injected at a given location, only the Poisson variance should be increased by $C_t \cdot\Delta \Ix$, while the background contribution $\varo[\Ix]$ remains unchanged leading to the variance update
\begin{equation}
\var[\Ix + \Delta \Ix] = \var[\Ix] + (C/t) \cdot\Delta \Ix \;.
\end{equation}

Since the uncertainty in the flux depends on the flux itself, the noise in the variance-flux relation is heteroscedastic. In particular, pixels with low flux have lower signal-to-noise ratio (S/N), making the observed flux a less reliable estimator of its expectation value and introducing larger scatter in the variance-flux plane. Therefore, we perform the linear fit using only the higher S/N region, when such pixels are sufficiently available. In practice, we restrict the fit to pixels with $\mathrm{S/N} > 5$, where the flux measurements more faithfully trace the underlying relation\footnote{
One can theoretically calculate the flux corresponding to a S/N=$R$,
    \begin{equation}
        I_{\text{S/N=}R} = \frac{R^2 ~(\text{slope})}{2} \left[1+\sqrt{1+\frac{4~(\text{intercept})}{R^2(\text{slope})^2}} ~\right]
    \end{equation}
}. 
    
Fig.~\ref{fig:var_fit_demo} illustrates the variance-flux relation for two representative observations (left and right panels), shown here using LRGs for demonstration, although the behaviour is generic. Each column corresponds to one example, with the top panels showing the image and its variance map, and the bottom panels showing the corresponding variance-flux relation.

In the left column (corresponding to the first example in Figs.~\ref{fig:host_light_demo} and \ref{fig:ts_demo}), the variance scales linearly with flux across the full dynamic range, and a single linear relation provides an excellent description, particularly for high-S/N pixels (S/N $> 5$). In the right column (corresponding to the third example in Fig.~\ref{fig:ts_demo}), the variance map exhibits spatially patchy background structure, producing multiple approximately linear sequences with similar slopes but different effective background levels. Nevertheless, the high-S/N pixels remain well described by a linear relation. Since the variance correction described above depends only on the slope of this relation, the Poisson scaling can still be robustly determined.

\subsection{PSF variation and the variance map}\label{sec:psf_var_map}

Let the underlying true sky brightness be $S(\vecx)$. An observed image under a PSF is given by
\begin{equation}
 I(\vecx) = (\mathrm{PSF} * S)(\vecx) + e(\vecx),
~ \langle I(\vecx) \rangle = (S * \mathrm{PSF})(\vecx) \;,   
\end{equation}
where $e(\vecx)$ denotes the corresponding noise realization and $*$ represents the convolution operation. Under the standard assumption that the noise is unbiased and uncorrelated between pixels,
\begin{equation}\label{eq:noise_var}
\langle e(\vecx) \rangle = 0, ~~
\langle e(\vecx)\,e(\vecx') \rangle
= \mathrm{Var}[I(\vecx)]\,\delta_{\vecx,\vecx'} \;.
\end{equation}
From this point onward, to avoid confusion between functions used in convolution and their
values at a given pixel, we denote by $\langle I\rangle$ and $\var[I]$ the expectation value and variance map of the image as functions of pixel position, such that $\langle I(\vecx)\rangle$ and $\var[I(\vecx)]$ denote their values at pixel $\vecx$. 

Let a normalized convolution kernel $k(\vecu)$, defined over pixel offsets $\vecu$, relate two PSFs as
\begin{equation}
\mathrm{PSF}_2(\vecx)
= \sum_{\vecu} k(\vecu)\,\mathrm{PSF}_1(\vecx-\vecu),
~
\sum_{\vecu} k(\vecu) = 1 ,
\end{equation}
which we write compactly as $\mathrm{PSF}_2 = k * \mathrm{PSF}_1$. Since
convolution with a normalized kernel is a smoothing operation, $\mathrm{PSF}_2$ must be broader than $\mathrm{PSF}_1$.\footnote{%
If $\mathrm{PSF}_2$ were narrower than $\mathrm{PSF}_1$, no such kernel exists and a
deconvolution would be required, which is not considered in this work.} Let $I_1(\vecx)$ and $I_2(\vecx)$ denote the observed images under these two PSFs. Using the associativity of convolution, the expectation value of $I_2(\vecx)$ can be written as
\begin{equation}\label{eq:E_Ix2}
\langle I_2(\vecx)\rangle
= (S * k * \mathrm{PSF}_1)(\vecx)
= (k * \langle I_1\rangle)(\vecx) \;.
\end{equation}

From Eq.~\eqref{eq:varp_Ix}, the Poisson variance of $I_2(\vecx)$ then follows as
\begin{align}
\varp[I_2(\vecx)]
&= \left(C/t\right)\,\langle I_2(\vecx)\rangle \\
&= \left(C/t\right)\,(k * \langle I_1\rangle)(\vecx) \\
&= (k * \varp[I_1])(\vecx) \;, \label{eq:varp_Ix2}
\end{align}
where the exposure time $t$ is assumed to be the same for both observations.

On the other hand, the background variance is assumed to be spatially uniform and to depend
only on the exposure time, but not on the PSF. Therefore,
\begin{equation}\label{eq:varo_2}
\varo[I_2(\vecx)] = \varo[I_1(\vecx)] = \mathrm{const}
= (k * \varo[I_1])(\vecx) \;,
\end{equation}
where the last equality follows from the normalization condition
$\sum_{\vecu} k(\vecu)=1$.

Substituting the expressions for the Poisson and background variance from Eqs.~\eqref{eq:varp_Ix2} and \eqref{eq:varo_2} into Eq.~\eqref{eq:tot_var}, the total variance map of $I_2(\vecx)$ can be written as
\begin{equation}\label{eq:var_I2}
\var[I_2(\vecx)] = (k * \var[I_1])(\vecx) \;.
\end{equation}

Equations~\eqref{eq:E_Ix2} and \eqref{eq:var_I2} thus describe how the expected image and its
variance map transform under a change of PSF. However, in practice one has access only to a
single noisy realization of the image, $I_1(\vecx)$, and its associated variance map
$\var[I_1(\vecx)]$, while the expectation value $\langle I_1(\vecx)\rangle$ is not directly
observable.

A practical approach is to convolve the noisy image $I_1(\vecx)$ with the kernel $k$,
\begin{align}
I_{1,\rm conv}(\vecx)
&\equiv (k * I_1)(\vecx) \label{eq:Iconv} \\
&= (k * \langle I_1\rangle)(\vecx) + (k * e_1)(\vecx) \nonumber \\
&= \langle I_2(\vecx)\rangle + e_{1,\rm conv}(\vecx) \;,
\end{align}
which reproduces the correct expected image. However, this procedure modifies the noise
properties: after convolution, the noise becomes correlated and its variance is suppressed.
The convolved noise field is given by
\begin{equation}
e_{1,\rm conv}(\vecx)
= (k * e_1)(\vecx)
= \sum_{\vecu} k(\vecu)\,e_1(\vecx-\vecu) \;.
\end{equation}

The variance map of the convolved image $I_{1,\rm conv}(\vecx)$ is therefore
\begin{align}
\var[I_{1,\rm conv}(\vecx)]
&= \var[e_{1,\rm conv}(\vecx)]
= \langle e_{1,\rm conv}^2(\vecx)\rangle \\
&= \sum_{\vecu,\vecv} k(\vecu)\,k(\vecv)\,
\langle e_1(\vecx-\vecu)\,e_1(\vecx-\vecv)\rangle
\label{eq:step1} \\
&= \sum_{\vecu} k^2(\vecu)\,\var[I_1(\vecx-\vecu)]
\label{eq:step2} \\
&= (k^2 * \var[I_1])(\vecx) \;, \label{eq:var_Iconv}
\end{align}
where Eq.~\eqref{eq:step2} follows from the assumption of pixel-uncorrelated noise
(Eq.~\eqref{eq:noise_var}), and $k^2(\vecu)$ denotes the element-wise square of the kernel.
Since typically
\begin{equation}\label{eq:k_k2}
\sum_{\vecu} k^2(\vecu) \ll \sum_{\vecu} k(\vecu) = 1 \;,
\end{equation}
the noise in $I_{1,\rm conv}(\vecx)$ is strongly suppressed.

To restore the correct variance, we add a residual noise realization,
\begin{align}
I_{2,\rm sim}(\vecx)
&= I_{1,\rm conv}(\vecx) + e_{\rm resi}(\vecx) \;, \label{eq:var_restore1} \\
e_{\rm resi}(\vecx)
&\sim \mathcal{N}\!\left(
0,\,
(k * \var[I_1])(\vecx) - (k^2 * \var[I_1])(\vecx)
\right) , \label{eq:var_restore2}
\end{align}
such that $I_{2,\rm sim}(\vecx)$ matches both the expectation value and the variance of
$I_2(\vecx)$.

This procedure adds an uncorrelated noise component, given by Eq. \eqref{eq:var_restore2}, on top of the correlated noise
$e_{1,\rm conv}(\vecx)$ given by Eq. \eqref{eq:var_Iconv}. For a sufficiently broader target PSF$_2$, the kernel satisfies
Eq.~\eqref{eq:k_k2}, implying $\var[e_{1,\rm conv}(\vecx)]/ \var[e_{\rm resi}(\vecx)]\ll 1$ and that
the correlated noise is strongly suppressed by the added uncorrelated component. In the
opposite limit, $\mathrm{PSF}_2 \rightarrow \mathrm{PSF}_1$, Eq.~\eqref{eq:k_k2} no longer
applies and the kernel approaches a Kronecker delta,
$k(\vecu) \rightarrow \delta_{\vecu,\mathbf{0}}$, such that
$e_{1,\rm conv}(\vecx) \rightarrow e_{1}(\vecx)$ which is uncorrelated to begin
with. In this case, the residual noise term $e_{2,\rm resi}(\vecx)$ vanishes and the
procedure reduces to the original noise realization.

To ensure that the overall noise, including the Poisson component, varies consistently
across epochs with different PSFs, we nevertheless add an additional noise layer sampled
from the target variance map $\var[I_2(\vecx)]$ given in Eq.~\eqref{eq:var_I2}. This step suppresses any residual correlated noise and enforces the correct Poisson variance, at the cost of a
conservative reduction in image S/N by a factor of $\sqrt{2}$. 

Moreover, the use of a Gaussian distribution for $e_{2,\rm resi}$ in Eq. \eqref{eq:var_restore2} is appropriate, since the
background noise is assumed to be Gaussian and Poisson noise approaches Gaussianity in the
high-count regime.

In summary, given an observed image $I_1(\vecx)$, its variance map $\var[I_1](\vecx)$, and
PSF $\mathrm{PSF}_1$, Eqs.~\eqref{eq:Iconv} and
\eqref{eq:var_restore1}--\eqref{eq:var_restore2} provide a practical method to generate a
simulated image $I_{2,\rm sim}(\vecx)$ for any target PSF broader than the original one.
The corresponding variance map is given by Eq.~\eqref{eq:var_I2}. Again, we remind that using kernel convolution we can simulate observations with PSFs broader than the original one. 

\section{Implemention of SALT2-extended SN Ia lightcurves with intrinsic stretch and color}
\label{ref:SALT2}
\begin{figure*}[h!]
    \centering
    \includegraphics[width=\textwidth]{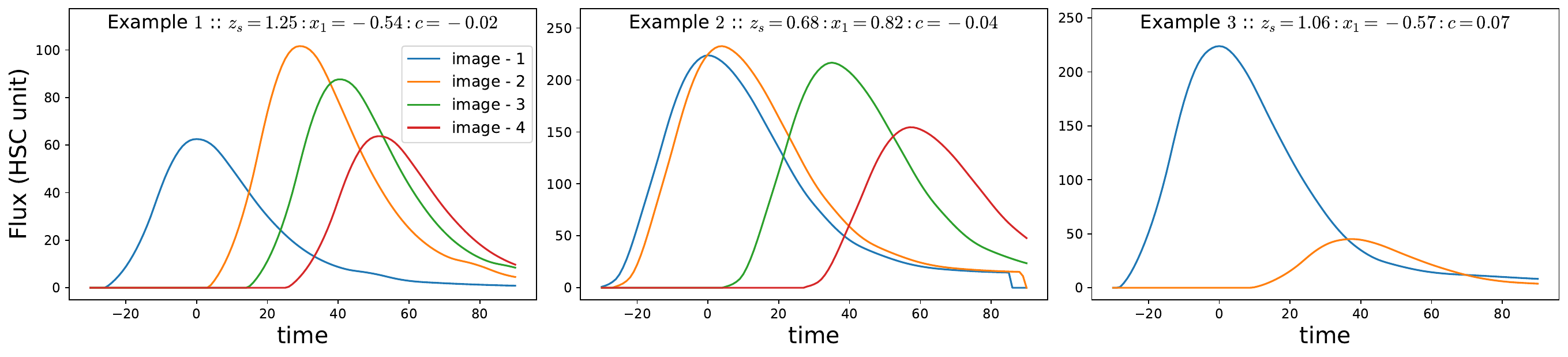}
\caption{Light curves of the LSNe Ia images in $i$-band, including the effects of lensing magnification and time delays for the lens configurations of three example systems, shown in the three panels. The two left panels correspond to the systems shown in Fig. \ref{fig:host_light_demo}. Time is measured relative to the 
$i$-band peak of the first-arriving image. These light curves determine the instantaneous brightness of the individual SN images and are used when painting the LSN images at the pixel level in the simulated images for these three systems shown in Fig. \ref{fig:ts_demo}. The stretch ($x_1$) and color ($c$) parameters used for these examples are sampled from the Pantheon+ distributions.
}
    \label{fig:lc_demo}
\end{figure*}

We model SNe Ia light curves using the SALT2 empirical model \citep{salt2}, in which the rest-frame spectral flux density is expressed as
\begin{equation}
F(t,\lambda) = x_0 \left[ \mathcal{M}_0(t,\lambda) + x_1,\mathcal{M}_1(t,\lambda) \right]
\exp\left[c,CL(\lambda)\right] ,
\end{equation}
where $t$ denotes the rest-frame phase relative to peak brightness and $\lambda$ is the rest-frame wavelength. The component $\mathcal{M}_0(t,\lambda)$ represents the globally trained mean spectral-temporal model of the SN Ia population, while $\mathcal{M}_1(t,\lambda)$ captures the leading variability associated with light-curve shape. The parameter $x_1$ describes the stretch of the light curve, $c$ represents the color parameter, $CL(\lambda)$ is the SALT2 color law, and $x_0$ sets the overall flux normalization.

Observed SN Ia magnitudes are commonly standardized using the Tripp relation \citep{Tripp1998},
\begin{equation}
M_B = M_{B,0} - \alpha x_1 + \beta c ,
\end{equation}
which accounts for the empirical correlations between luminosity, light-curve stretch, and intrinsic color. Here $M_{B,0}$ is the fiducial standardized absolute magnitude in the rest-frame $B$ band, while $\alpha$ and $\beta$ quantify the stretch-luminosity and color-luminosity relations.

For our simulations, we adopt $M_{B,0} = -19.3$ in the rest-frame Bessell-$B$ band. Light curves are generated using the \texttt{sncosmo} package \citep{sncosmo} with the \texttt{salt2-extended} spectral model, which extends the original \texttt{salt2} spectral coverage into the ultraviolet and near-infrared. The model amplitude is normalized using the rest-frame $B$-band absolute magnitude, after which \texttt{sncosmo} generates observer-frame light curves in the LSST filters by redshifting the spectral time series and integrating it through the filter transmission curves, thereby naturally incorporating cosmological dimming, time dilation, and cross-filter K-corrections.

For each SN Ia in our sample at a given redshift, we generate light curves by sampling $(x_1, c)$ from the empirical Pantheon+ distributions \citep{PantheonP}. The nuisance parameters $(\alpha, \beta)$ are drawn from the Pantheon+ posterior constraints assuming a $\Lambda$CDM cosmology, ensuring consistency with the Pantheon+ light-curve standardization.

Figure~\ref{fig:lc_demo} shows the $i$-band light curves of the SN images for three example LSNe~Ia systems, including the effects of lensing magnification and time delays. These curves determine the instantaneous brightness used when painting the SN images at the pixel level in the simulated time series shown in Fig.~\ref{fig:ts_demo}. The sampled stretch ($x_1$) and color ($c$) values used for these examples are indicated in the corresponding panels.

\section{Potential confusion with sibling supernovae}
\label{sec:sibling}

\begin{figure}[h!]
    \centering
    \includegraphics[width=\linewidth]{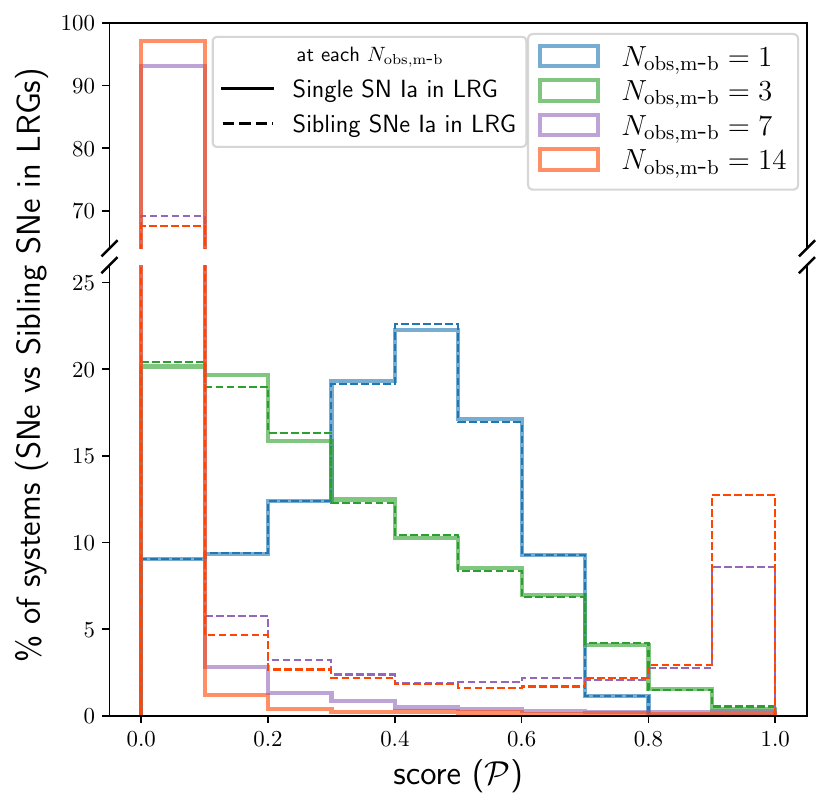}
    \includegraphics[width=\linewidth]{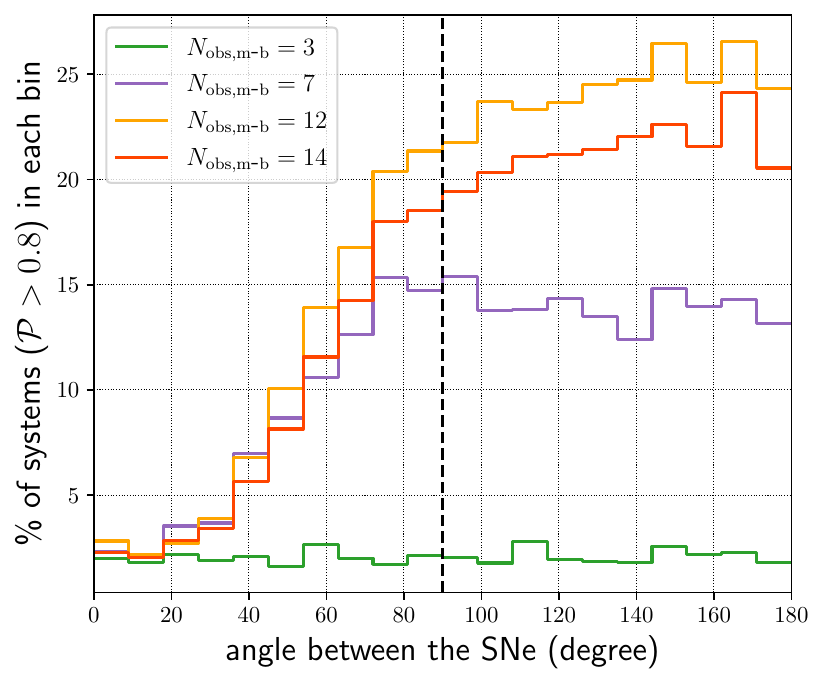}
    \caption{Classification behaviour of the model for sibling SNe, where two independent SNe Ia occur in the same host galaxy within a short time window and are simultaneously visible. Such systems are rare but can geometrically resemble lensed double-image configurations and were not included in the training set.
\textbf{Top:} Distribution of model classification scores for sibling SNe (dashed histograms) compared with the primary negative sample ``SN in LRGs'' (solid histograms) after different numbers of observing epochs ($\Ne=1,3,7,14$). While the two populations are indistinguishable at early epochs, a fraction of sibling SNe receive high scores at later epochs once both explosions are visible.
\textbf{Bottom:} Fraction of sibling SNe receiving high scores ($\mathcal{P}>0.8$) as a function of the projected angle between the two SNe with respect to the host galaxy centre. The fraction increases with angular separation and saturates at $\gtrsim90^\circ$, indicating that configurations where the two SNe lie on opposite sides of the galaxy more closely resemble double-image lens geometries. 
}
    \label{fig:sibling_sne_dist}
\end{figure}

Sibling SNe, defined as multiple SNe occurring in the same galaxy, can represent a potential source of confusion for LSNe searches when two such events occur within a sufficiently short time interval that both are visible simultaneously \citep{Stritzinger2002}. In these situations the two SNe may visually resemble multiply imaged transients. At the same time, sibling SNe are themselves scientifically interesting events, as they provide opportunities to study SN progenitor diversity, test SN Ia standardization and intrinsic scatter, and investigate host-galaxy environmental effects, since the events occur in the same galaxy and therefore share the same distance and global environment \citep{brown2014, Scolnic2020, Kelsey2023}.

Although the rate of such simultaneously visible sibling SNe is expected to be low, it is not negligible, and it is not immediately clear whether these events or lensed SNe are intrinsically rarer. It is therefore useful to examine how our model responds to such configurations. Because sibling SNe constitute rare and scientifically valuable transients, we deliberately excluded them from the training process. Occasional classification of these systems as lensed candidates would therefore not be problematic, provided they do not dominate the false-positive population. Here we evaluate the model response on a dedicated sample of such systems.

To construct the test set, we inject two SNe Ia into LRGs using the same procedure adopted for the primary negative subset ``SN in LRGs'' (set $N_3$). The two SNe are assigned an explosion-time offset drawn from $t_{\rm offset} \sim \mathcal{U}(0,30)~{\rm days}$. Their positions are sampled independently within $\sim \mathcal{U}(0,2R_{50})$, where $R_{50}$ is the half-light radius of the LRG, with the separation clipped at $2''$, and their relative orientation is unrestricted. Each SN is assigned independent colour and stretch parameters drawn from the Pantheon+ population. The first detection epoch is defined by the arrival of the earlier SN, as in the case of set $N_3$.

Figure~\ref{fig:sibling_sne_dist} presents the classification results. The top panel shows the predicted score distributions after four observing epochs ($\Ne = 1, 3, 7, 14$). For each epoch, the dashed histogram represents the sibling SNe sample, while the solid histogram shows the ``SN in LRGs'' sample for comparison. At early epochs the two populations exhibit little difference, as the second SN has not yet appeared in most sibling systems. At later epochs ($\Ne = 7, 14$), however, an excess of sibling SNe emerges at higher score values. Approximately $10$--$15\%$ of sibling SNe receive scores $>0.9$ at $\Ne = 7$ and $14$, whereas only $\mathcal{O}(0.01\%)$ of the single-SN systems reach such high scores. Thus the model flags a non-negligible fraction of these systems as lensed. Nevertheless, the majority ($\gtrsim 85\%$, depending on the observing epoch) still receive low scores and are correctly classified as unlensed, likely due to colour information, brightness ratios, or inconsistencies with plausible lensing configurations.
For comparison, more than $80$--$90\%$ of genuine doubly lensed SNe Ia receive scores above $0.9$ by the $7$th and $14$th observations, respectively.

This behaviour is not unexpected. In particular, when two sibling SNe occur on opposite sides of the host galaxy they can visually resemble double-image lens configurations. To investigate this further, the bottom panel of Fig.~\ref{fig:sibling_sne_dist} shows the fraction of systems receiving high scores ($P>0.8$) as a function of the angle between the two SNe projected with respect to the galaxy centre. After the first observation no system received such high score, therefore that epoch is not shown. At the third observation there is still little dependence on angle. However, at later epochs the fraction of high-score systems increases with separation angle, rising beyond $\theta \gtrsim 45^\circ$, saturating near $\theta \sim 90^\circ$, and reaching approximately $\sim 15\%$, $25\%$, and $22\%$ for $\Ne = 7,12$, and $14$, respectively. The increase even for $\theta < 90^\circ$ suggests that some of these systems resemble quadrupole lensed configurations such as cusp or fold image pairs, particularly when earlier-arriving images, if any, in those configurations are missed. Another interesting feature is that the fraction of high-scoring systems peaks at $\Ne = 12$ and then decreases at higher $\Ne$, hinting that as more of the light curve becomes available, the model begins to distinguish the sibling SNe systems more effectively from genuine lensed cases.

Nevertheless, in all angular bins the majority of systems still receive low classification scores. This indicates that the model is generally able to correctly identify them as unlensed events even though such systems were never included during training. The model likely relies on additional cues such as colour evolution, brightness ratios, and inconsistencies with realistic lensing geometries.

\end{document}